%% file: sample-sigconf.tex
\documentclass[sigconf]{acmart}
\usepackage{enumitem}
\usepackage{graphicx}
\usepackage{wrapfig}
\usepackage{makecell}
\usepackage[table]{xcolor}   
\usepackage{tcolorbox}       
\usepackage{url}
\usepackage{booktabs}
\usepackage{multirow}
\usepackage{xcolor}  
\definecolor{lightgray}{gray}{0.3}
\usepackage{seqsplit}
\usepackage{graphicx}    
\usepackage{subcaption}  
\usepackage{mathrsfs}
\usepackage{subcaption}
\usepackage{algorithm}
\usepackage{algpseudocode}
\usepackage{colortbl}
\usepackage{balance}

\definecolor{mygreen}{HTML}{228B22}

\AtBeginDocument{%
  }

\copyrightyear{2026}
\acmYear{2026}
\setcopyright{cc}
\setcctype{by}
\acmConference[KDD '26]{Proceedings of the 32nd ACM SIGKDD Conference on Knowledge Discovery and Data Mining V.2}{August 09--13, 2026}{Jeju Island, Republic of Korea}
\acmBooktitle{Proceedings of the 32nd ACM SIGKDD Conference on Knowledge Discovery and Data Mining V.2 (KDD '26), August 09--13, 2026, Jeju Island, Republic of Korea}
\acmDOI{10.1145/3770855.3817477}
\acmISBN{979-8-4007-2259-2/2026/08}

\begin{document}

\title{FORGE: Forming Semantic Identifiers for Generative Retrieval in Industrial Datasets}

\author{Kairui Fu}
\authornote{Equal contribution. Work is done during Kairui Fu's internship at Alibaba Group.}
\affiliation{%
  \institution{Zhejiang University}
  \city{Hangzhou}
  \country{China}}
\email{fukairui.fkr@zju.edu.cn}

\author{Tao Zhang}
\authornotemark[1]
\affiliation{%
  \institution{Taobao \& Tmall Group of Alibaba}
  \city{Hangzhou}
  \country{China}}
\email{selous.zt@alibaba-inc.com}

\author{Shuwen Xiao}
\authornotemark[1]
\affiliation{%
  \institution{Taobao \& Tmall Group of Alibaba}
  \city{Hangzhou}
  \country{China}}
\email{shuwen.xsw@alibaba-inc.com}

\author{Ziyang Wang}
\affiliation{%
  \institution{Taobao \& Tmall Group of Alibaba}
  \city{Hangzhou}
  \country{China}}
\email{shanyi.wzy@alibaba-inc.com}

\author{Xinming Zhang}
\affiliation{%
  \institution{Taobao \& Tmall Group of Alibaba}
  \city{Hangzhou}
  \country{China}}
\email{zxm454149@taobao.com}

\author{Chenchi Zhang}
\affiliation{%
\institution{Taobao \& Tmall Group of Alibaba}
\city{Hangzhou}
\country{China}
}
\email{zhangchenchi.zcc@taobao.com}

\author{Yuliang Yan}
\authornote{Corresponding authors.}
\affiliation{%
\institution{Taobao \& Tmall Group of Alibaba}
\city{Hangzhou}
\country{China}
}
\email{yuliang.yyl@alibaba-inc.com}

\author{Junjun Zheng}
\affiliation{%
\institution{Taobao \& Tmall Group of Alibaba}
\city{Hangzhou}
\country{China}
}
\email{fangcheng.zjj@alibaba-inc.com}

\author{Xiangheng Kong}
\affiliation{%
\institution{Taobao \& Tmall Group of Alibaba}
\city{Hangzhou}
\country{China}
}
\email{yongheng.kxh@alibaba-inc.com}

\author{Shengyu Zhang}
\affiliation{%
\institution{Zhejiang University}
\city{Hangzhou}
\country{China}
}
\email{sy_zhang@zju.edu.cn}

\author{Kun Kuang}
\authornotemark[2]
\affiliation{%
\institution{Zhejiang University}
\city{Hangzhou}
\country{China}
}
\email{kunkuang@zju.edu.cn}

\author{Yuning Jiang}
\affiliation{%
\institution{Taobao \& Tmall Group of Alibaba}
\city{Beijing}
\country{China}
}
\email{mengzhu.jyn@alibaba-inc.com}

\renewcommand{\shortauthors}{K Fu et al.}

\begin{abstract}
    Semantic identifiers (SIDs) have gained increasing attention in generative retrieval (GR) for recommendation due to their meaningful semantic discriminability. However, current studies in this field primarily (1) offer limited investigation into the construction strategies for better SIDs, and (2) their SID assessment typically relies on costly GR training. To address these challenges, we propose \textbf{FORGE}\raisebox{-0.015in}{\includegraphics[height=0.14in, width=0.17in]{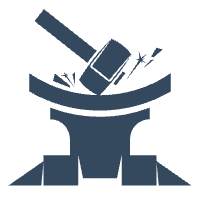}}, a comprehensive benchmark for \textbf{FO}rming semantic identifie\textbf{R}s for \textbf{G}enerative r\textbf{E}trieval. Specifically, FORGE provides a taxonomy of the SID construction process from several perspectives and validates their impact on downstream GR through offline experiments across diverse settings. Notably, these empirical findings have led to a 0.35\% increase in transaction count via online A/B experiments in the Guess You Like section of Taobao. The corresponding SID construction strategies have since been deployed at full scale on Taobao, demonstrating their practical effectiveness. To avoid expensive SID assessment that requires full GR training, we propose two novel SID evaluation metrics that are highly correlated with recommendation performance, enabling convenient evaluations without any GR training. Furthermore, to facilitate the community, we release AL-GR, the industrial dataset used in our experiments, comprising 14 billion interactions and 250 million items with the corresponding multimodal features collected from Taobao. All the code and data are available at \textcolor{blue}{\url{https://github.com/selous123/al_sid}}.
\end{abstract}



\maketitle

\input{./files/introduction}

\input{./files/related-work}

\input{./files/framework}

\input{./files/experiment}

\input{./files/conclusion}

\begin{acks}
This work was supported in part by "Pioneer" and "Leading Goose" R\&D Program of Zhejiang (2025C02037), and National Natural Science Foundation of China (62376243). All opinions in this paper are those of the authors and do not necessarily reflect the views of the funding agencies.
\end{acks}

\clearpage

\bibliographystyle{ACM-Reference-Format}
\balance
\bibliography{sample-base}

\input{files/appendix}

\end{document}

%% file: files/introduction.tex
\section{Introduction}
With the ability to predict the next item in an end-to-end manner, generative retrieval (GR) has recently emerged as a promising approach in recommender systems~\cite{rajput2023recommender,liang2025tbgrecall}. Typically, this framework encodes user behaviors into a sequence of \textit{identifiers}, employs larger models to capture item dependencies, and generates \textit{identifiers} of candidates as results. Therefore, the tokenization, which determines identifiers for each item, plays a fundamental role in advancing accurate recommendation~\cite{lin2025order}.

\begin{figure}[htb]
    \centering
    \includegraphics[width=0.88\columnwidth]{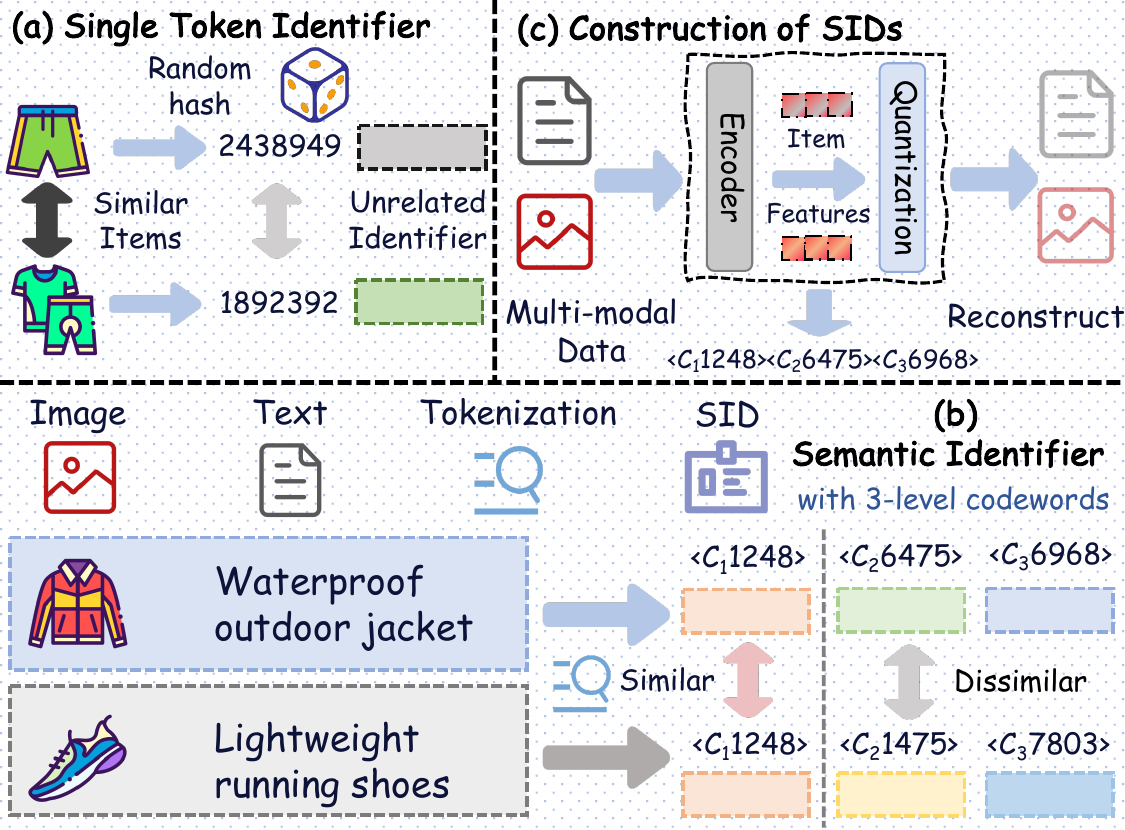}
    \vspace{-0.20cm}
    \caption{Diagram of (a) single token identifier and (b) semantic identifier. (c) The construction of semantic identifiers.}
    \vspace{-0.55cm}
    \label{fig: diagram}
\end{figure}
Recent advancements in item tokenization for recommendation can be broadly categorized into single-token identifiers and semantic identifiers (SIDs). 
\textbf{\romannumeral1)} In Figure~\ref{fig: diagram}(a), the former randomly assigns a unique identifier to each item (e.g., 1892392$\rightarrow$\raisebox{-0.05in}{\includegraphics[height=0.16in, width=0.18in]{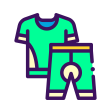}}) and updates the corresponding latent token independently during training. This paradigm requires managing massive vocabularies with billions of items in real-world recommender systems~\cite{barkan2016item2vec}. Numerous identifiers prevent it from including all items, but only several negative ones are sampled for loss calculation to reduce computational burdens~\cite{zhai2024actions,ma2024negative}.
Such an approximation, however, might result in inconsistency between offline training and online serving, where the latter requires measuring all items.
\textbf{\romannumeral2)} The semantic identifiers shown in Figure~\ref{fig: diagram}(b) represent each item with multi-level \textbf{codewords} (e.g., <$C_{1}1248$><$C_{2}6475$><$C_{3}6968$>$\rightarrow$\raisebox{-0.045in}{\includegraphics[height=0.16in, width=0.18in]{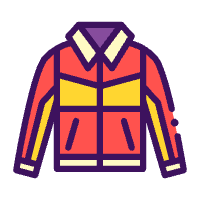}}), and recommenders would generate the top-K item identifiers as candidates with continuous beam search~\cite{li2025bbqrec,zheng2025universal}. The codewords in an SID are constructed from multimodal data, thus enabling knowledge sharing among similar items. Furthermore, with hierarchical codewords, SIDs could compactly encode an exponentially large item space (e.g., a 3-level SID with 8192 tokens per level could represent $8192^3 \gg 10^{10}$ items using 24576 tokens). The relatively small number of tokens also makes it computationally feasible to include all of them in the calculations during training~\cite{rajput2023recommender}.

As depicted in Figure~\ref{fig: diagram}(c), the construction process for SIDs typically follows an encoder-quantization framework: an encoder for extracting semantic features from raw multimodal data and another quantization module~\cite{esser2021taming,lee2022autoregressive} for SID encoding, separately~\cite{zheng2024adapting}. These compact yet semantically rich identifiers subsequently remain unchanged and are used by downstream retrieval models during task-specific training. Despite their promising results on real-world datasets~\cite{liu2025generative,ju2025generative}, they still have the following challenges that hinder the development of this field: \textbf{\romannumeral1) Neglect of strategy selection for SID construction}. Even if SID serves as the foundation for the retrieval task~\cite{yang2024unifying,ju2025generative}, useful insights around the design choices for it are typically not well-discussed in previous literature, making it complicated for researchers to select an optimal configuration. \textbf{\romannumeral2) Inefficient enumeration for a better SID configuration}. Recent works assess the quality of SIDs through time-consuming GR training~\cite{ju2025generative}, which is computationally expensive for large-scale applications. This highlights a critical need for direct metrics with convenient evaluations of SIDs.

To mitigate this gap, we propose \textbf{FORGE}\raisebox{-0.01in}{\includegraphics[height=0.14in, width=0.17in]{./figs/title.png}}, a comprehensive benchmark for \textbf{FO}rming semantic identifie\textbf{R}s for \textbf{G}enerative r\textbf{E}trieval in industrial datasets. 
(\textbf{Challenge \romannumeral1}) FORGE investigates several strategies related to SID construction from multiple angles, encompassing diverse data modalities, SID encoding, and ID collision mitigation (i.e., cases where multiple similar items correspond to the same SID). We then thoroughly validate the influence of these strategies across a range of settings, including different downstream GR models, SID structures, and search tasks. Notably, these discovered insights also bring several improvements in the 7-day online evaluation on Taobao, thus supporting the validity of our benchmark. (\textbf{Challenge \romannumeral2}) To facilitate the SID evaluation, we introduce two novel metrics (i.e., \textit{Embedding HitRate} and \textit{Gini coefficient}) to assess SID quality directly. Experiments show that these metrics exhibit strong correlation with GR performance, enabling SID measurements without costly GR training and thus providing an effective and efficient evaluation process. Additionally, we also open-source the industrial-scale dataset used in the experiments, which contains more than 14 billion user behaviors in 10 days, along with the multimodal features of 250 million items collected from Taobao, one of the biggest e-commerce platforms in China. We summarize the major contributions as follows:
\begin{enumerate}[leftmargin=*]
  \item To identify which configurations are most beneficial for SID construction, we propose FORGE, which systematically investigates strategies from multiple perspectives, including data modality, encoding schemes, and collision mitigation.
  \item For the expensive SID evaluation, which requires training the entire GR model, we propose two novel metrics whose trends are strongly aligned with GR performance, enabling cost-free SID quality assessment without any GR training. 
  \item We conduct extensive offline experiments under different settings and yield several practical insights. The subsequent 0.35\% improvement in transaction count in online A/B experiments of Taobao further supports the practical value of our findings.
  \item We collect and release the AL-GR dataset, which is built from industrial-scale user behaviors with rich multimodal features in Taobao. To the best of our knowledge, it is the first industrial-scale dataset for generative retrieval with SIDs.
\end{enumerate}

%% file: files/related-work.tex
\section{Related Work}
\subsection{Generative Retrieval} The rapid development of recommender systems has drawn a surge of research attention in recent years~\cite{bobadilla2013recommender,zhang2019deep,yu2025thinkrec,fu2024diet,fu2023end,fu2025forward}.
As the initial filtering component in recommendation, the retrieval stage quickly narrows down the item pool from billions to tens of thousands of interested candidates. Traditional retrieval methods~\cite{kang2018self,hidasi2015session,li2019multi} encode users and candidates into dense embeddings with a unified space, followed by Approximate Nearest Neighbors to select several related items given each user. 
After KE~\cite{de2021editing} introduces the notion of editing and retrieving via generation, generative retrieval~\cite{rajput2023recommender,deng2025onerec,zheng2025ega} becomes an innovative end-to-end approach for both recommendation and search, which directly decodes the identifiers of user-interested items in an auto-regressive way. Based on this, subsequent work adapted more suitable training frameworks~\cite{liang2025tbgrecall,liu2025generative} and introduced strategies like reinforcement learning~\cite{deng2025onerec,zheng2025ega} to generate items that better align with user preferences.

\begin{figure*}[ht]
    \centering
    \includegraphics[width=1.0\linewidth]{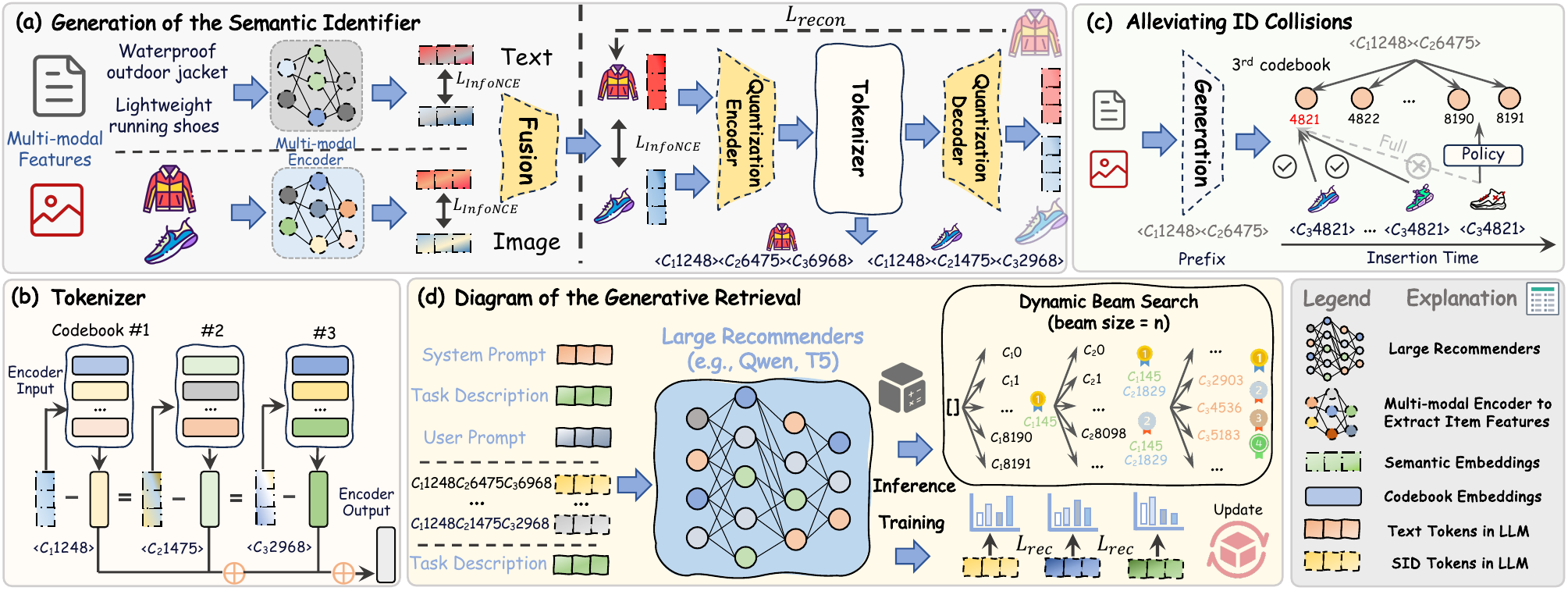}
    \vspace{-0.5cm}
    \caption{Overview of GRs. (a) The generation process of SIDs. Left panel: the training of the text/image encoder\raisebox{-0.05in}{\includegraphics[height=0.18in, width=0.18in]{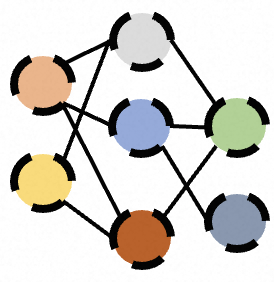}}. Right panel: the schematic diagram of item tokenization. (b) Detailed workflow of the tokenizer. (c) Post-collision handling for semantic identifiers. (d) The training and dynamic inference procedure of generative retrieval.}
    \vspace{-0.3cm}
    \label{fig: intro}
\end{figure*}

\subsection{Semantic Identifier} The innovative SID widely adopted in both search~\cite{sun2023learning,zeng2024scalable} and recommendation~\cite{rajput2023recommender} utilizes multiple codewords to build identifiers for items, enabling knowledge sharing among similar ones. Subsequent researchers~\cite{zhang2025replication,wu2025constrained} investigated the generalization ability of generative retrieval when facing unseen new items. The widespread use in two domains has also encouraged some work~\cite{penha2024bridging,penha2025semantic,zhao2025unifying} that focuses on generalizing SIDs well in a joint setting. From the perspective of modality to build SIDs, several works~\cite{wang2024content,liu2025generative,wang2024learnable,li2025bbqrec} incorporate additional knowledge from user-item interactions to align the multimodal features of items, thereby equipping the generated SIDs with both semantic and collaborative information in recommendation. As for the encoding, OneRec~\cite{deng2025onerec} and RPG~\cite{hou2025generating} tend to simplify the quantization process and accelerate the decoding process via RQ-Kmeans and optimized product quantization (OPQ)~\cite{ge2013optimized}. Apart from these studies, FORGE further investigates the entire SID construction pipeline from a holistic perspective, covering feature extraction, SID encoding, and postprocessing. Such a systematic investigation enables FORGE to offer practical guidance on designing better SIDs for the community.

%% file: files/framework.tex
\section{Benchmark Design}
\label{Framework}
This section provides an overview of our benchmark. Specifically, we first introduce the fundamental steps of SID construction in Section~\ref{SID Construction}, which mainly focuses on three key aspects: feature extraction, SID encoding, and SID post-processing. Sections~\ref{Generative Retrieval} and~\ref{Datasets} then describe the GR models and datasets used in our experiments, with particular emphasis on our industrial-scale dataset collected from Taobao. Finally, we introduce the evaluation metrics for both SID quality and recommendation performance in Section~\ref{Evaluation Metrics}.

\subsection{SID Construction}
\label{SID Construction}

\subsubsection{Feature Extraction}
\label{feature extraction}

Capturing the underlying item semantics through the multimodal content is a key step in generating SIDs. A natural approach is to leverage pretrained multi-modal large language models like CN-CLIP~\cite{yang2022chinese}, which consists of $\mathcal{M}_{\text{text}}$ and $\mathcal{M}_{\text{image}}$ in Figure~\ref{fig: intro}(a), to extract and fuse features $\mathcal{H}^{i}$ from text $\mathcal{I}^i_{\text{text}}$ and image $I^i_{\text{image}}$ of item $i$:
\begin{equation}
    \mathcal{H}^{i}=\mathcal{M}_\text{Fusion}(\mathcal{M}_{\text{text}}(\mathcal{I}^{i}_{text}), \mathcal{M}_{\text{image}}(\mathcal{I}^{i}_{image})).
\end{equation}
The $\mathcal{M}_{\text{Fusion}}$ module utilizes QFormer~\cite{jiang2024qformer} to integrate information from two modalities.
To further enhance the discriminability of these representations and reflect patterns from user behaviors, we could employ another contrastive loss $L_{\text{InfoNCE}}$. It encourages items that frequently co-occur to have more similar representations:
\begin{equation}
\label{contrastive}
\begin{split}
L_{\text{InfoNCE}} =  f(\mathcal{H}^i_{\text{text}}, \mathcal{H}^{i^+}_{\text{text}}, \mathcal{H}^{i^-}_{\text{text}}) & +
f(\mathcal{H}^i_{\text{image}}, \mathcal{H}^{i^+}_{\text{image}}, \mathcal{H}^{i^-}_{\text{image}}) \\
& + f(\mathcal{H}^i, \mathcal{H}^{i^+}, \mathcal{H}^{i^-})
\end{split}
\end{equation}
where $f$ denotes the InfoNCE loss~\cite{radford2021learning}. The positive $i^{+}$ and negative $i^{-}$ samples are collected from the i2i table (i.e., table recording the co-occurrence counts of items) and in-batch sampling, respectively.

\subsubsection{SID Encoding}
The multi-modal feature $\mathcal{H}^i$ of item $i$ is then used to tokenize its SID. To ensure compatibility with the supervised learning framework and the autoregressive nature of generative retrieval, several methods select the RQ-VAE~\cite{lee2022autoregressive} in Figure~\ref{fig: intro}(b) to convert $\mathcal{H}^i$ into a sequence of discrete codewords $\{c_1, ..., c_m\}$, with $m$ as the number of codewords. Formally, the encoding process of the $l$-th codeword $c_l$ can be formulated as:
\begin{equation}
    c_l = {\arg\min}_{c \in [1, n_l]} \left\| h_l - q_l^c \right\|,
\end{equation}
where $h_l$ is obtained as the residual between the input feature $h_{l-1}$ and the nearest vector $q_{l-1}^{c_{l-1}}$ from the previous codebook. At the first level ($l=1$), the initial residual is defined as $h_{1}=\mathcal{H}^{i}$. In the experiment, we compare the RQ-VAE with alternatives like random assignment, multiple VQ~\cite{esser2021taming}, Optimized Product Quantization (OPQ)~\cite{hou2025generating}, and RQ-Kmeans~\cite{deng2025onerec} to validate the effectiveness.

\subsubsection{SID post-processing}
\label{ID collisions}
During the SID generation, multiple similar items from different configurations or shops may be assigned the same SID, a situation referred to as \textbf{ID collision}. This may also result in low utilization, where certain SIDs remain unassociated with any item. To alleviate this issue, in Figure~\ref{fig: intro}(c), we introduce two optional post-processing strategies to control the maximum number of items mapped to each SID. The \textbf{KNN-based} strategy samples multiple candidate SIDs and evaluates them sequentially based on their associated scores until a SID with fewer than $\sigma$ corresponding items is found, where $\sigma$ is a predefined threshold.
In contrast, the \textbf{random-based} strategy focuses on dispersing the SID distribution at the final level rather than preserving semantic consistency. It assigns incremental codebooks to the last level in a circular manner, following the insertion order of items sharing the same prefix.
The pseudocode of the KNN-based policy and the random-based policy are shown in Algorithm~\ref{alg:KNN} and Algorithm~\ref{alg:Random}, respectively.
Both approaches lead to a fairer distribution of codebook usage and further improve the performance of GR.

\vspace{-0.2cm}
\begin{algorithm}[htb]
\caption{KNN-based Policy for alleviating ID Collisions}
\label{alg:KNN}
\centering
\begin{algorithmic}[1]
    \State $\text{codeword1}, \dots, \text{\{multiple\_codewords3\}} \gets \Call{Encoding}{\mathcal{H}^i}$
    \ForAll{$\text{codeword3} \in \text{\{multiple\_codewords3\}}$}
        \State $\text{SID} \gets [\text{codeword1}][\text{codeword2}][\text{codeword3}]$
        \If{$\text{count\_items}(\mathrm{SID})< \text{predefined threshold}\ \sigma$}
            \State \Return SID
        \EndIf
    \EndFor
\end{algorithmic}
\end{algorithm}
\vspace{-0.5cm}
\begin{algorithm}[htb]
\caption{Random-based Policy for alleviating ID Collisions}
\label{alg:Random}
\centering
\begin{algorithmic}[1]
    \State $\textbf{\text{Global}}\ \mathrm{IndexMap} \gets \{\}$  \Comment{key=(c1,c2), value=index}
    \State $\text{codeword1}, \text{codeword2} \gets \Call{Quantization}{\mathcal{H}^i}$
    \State $\text{key} \gets (\text{codeword1}, \text{codeword2})$
    
    \If{$\text{key} \notin \mathrm{IndexMap}$}
        \State $\mathrm{IndexMap}[\text{key}] \gets 0$  \Comment{Initialize index to $0$ on first occurrence}
    \EndIf
    
    \State $\text{codeword3} \gets \mathrm{IndexMap}[\text{key}]$
    \State $\mathrm{IndexMap}[\text{key}] \gets (\mathrm{IndexMap}[\text{key}] + 1) \bmod n_3$ \Comment{$n_3$ is the size of  third-level codebook}
    
    \State \Return $[\text{codeword1}][\text{codeword2}][\text{codeword3}]$
\end{algorithmic}
\end{algorithm}
\vspace{-0.5cm}

\subsection{Generative Retrieval}
\label{Generative Retrieval}
Once the set of SIDs is constructed, we concatenate the SIDs derived from user behavior with system instructions and user-specific prompts to form the input $x$ for GRs, which can be implemented with arbitrary architectures like Qwen~\cite{team2024qwen2} or T5~\cite{ni2022sentence}. Tokens of codewords are extended into the tokenizer $\tau$ and jointly trained with the recommender $\theta$. Formally, given a sequence of codewords $\{c_1, c_2, ...,c_m\}$ representing the SID of the target item, we use cross-entropy loss~\cite{mannor2005cross} to optimize both $\theta$ and $\tau$ by maximizing the conditional probability distribution $P_{\theta,\tau}(c_i|x,c_{j < i})$:
\begin{equation}
    \mathcal{L}_{rec}= - \sum_{i=1}^{m} \log P_{\theta,\tau}(c_i \mid x, c_{<i}).
\end{equation}
For inference, we employ beam search~\cite{freitag2017beam} and retain several SIDs as the retrieval results. However, maintaining a fixed beam width across all decoding steps can introduce significant computational overhead, particularly in platforms with ten thousand requests per second. Under these circumstances, each step must receive K sequences as input, resulting in a computational cost of K times. To address this, FORGE is equipped with \textbf{dynamic beam search} as illustrated in Figure~\ref{fig: intro}(d). Specifically, we set the beam size to 300 for the first, 600 for the second, and $K=$1200 for the final decoding step. Therefore, the first decoding step retains only 300 sequences, and the second decoding step processes significantly fewer sequences than the non-dynamic approach, which would involve 1200 sequences as input. As a result, the computational cost of the second decoding step is reduced by 75\%, and the number of SIDs remains 1200 after the third decoding step.

\begin{table}[ht]
\tabcolsep=4.5pt
\vspace{-0.1cm}
\caption{Statistics of the dataset.}
\vspace{-0.3cm}
\label{tab:dataset}
\begin{tabular}{@{}ccccc@{}}
\toprule
                      & \textbf{\#users} & \textbf{\#Items} & \textbf{\#Interactions} & \textbf{Sparsity}\\ \midrule
\textit{Amazon} & 2,524,981           & 714,957            & 12,530,806               & 99.98\%        \\
\textit{AL-GR}   & \textbf{131M}            & \textbf{251M}           & \textbf{14B}             & 99.99\%           \\
   
\bottomrule
\end{tabular}
\tabcolsep=7pt
\vspace{-0.3cm}
\end{table}

\begin{table*}[hbt]
\centering
\caption{Dataset statistics and properties. The proposed AL-GR contains a large-scale, temporally continuous dataset with multimodal content and a hierarchical codebook for semantic modeling.}
\vspace{-0.24cm}
\label{tab:dataset_stats}
\resizebox{0.80\textwidth}{!}{
\begin{tabular}{lcccccccl}
\toprule
\multirow{2}{*}{\textbf{Dataset}} & \textbf{Temporal} & \multirow{2}{*}{\textbf{\#Users}} & \multirow{2}{*}{\textbf{\#Items}} & \multirow{2}{*}{\textbf{\#Interactions}} & \multirow{2}{*}{\textbf{Modalities}} & \multirow{2}{*}{\textbf{Codebook}} & \multirow{2}{*}{\textbf{Tasks}} \\
                 & \textbf{Continuity} & & & & & & \\
\midrule
MovieLens & N & 138K & 27K & 20M & ID, text & N & Recsys \\
Yelp        & N & 2.1M & 160K & 8M & ID, text & N & Recsys \\
Taobao      & N & 987K & 4M & 100M & ID & N & Recsys \\
TenRec      & N & 5.0M & 3.7M & 142M & ID & N & Recsys \\
KuaiRec     & N & 7K & 10K & 12M & ID, tags & N & Recsys \\
RecFlow     & N & 42K & 82M & 38M & ID & N & Recsys \\
\midrule
\textbf{AL-GR} & \textbf{Y} & \textbf{131M} & \textbf{251M} & \textbf{14B} & \textbf{ID, text, image} & \textbf{Y} & \textbf{Recsys, Search} \\
\bottomrule
\vspace{-0.6cm}
\end{tabular}
}
\end{table*}

\subsection{Datasets}
\label{Datasets}

To enhance the reliability of our benchmark, we carry out experiments on two datasets with different scales and sparsity, whose descriptions can be found in Table~\ref{tab:dataset}. Notably, apart from the widely adopted Amazon-Clothing\footnote{\url{https://amazon-reviews-2023.github.io/}} dataset with relatively sparse characteristics in research, to bridge the gap between academic research and real-world applications and enable insights derived from FORGE to be validated in industrial settings, we collected and released\footnote{\url{https://huggingface.co/AL-GR}} the AL-GR dataset from Taobao, one of the largest e-commerce platforms in China. 

\subsubsection{The Industrial AL-GR Dataset}
To the best of our knowledge, AL-GR is the first industrial-scale dataset specifically for generative retrieval. Specifically, it contains two main types:

\textbf{\romannumeral1)} \texttt{Seq Data} samples 40 million user behavior records per day from Taobao over ten consecutive days for training. To enhance reliability, FORGE supports another dataset for the search task, whose data follows the structure of the recommendation, except for the inclusion of an extra query keyword. The data is divided into three stages (S1, S2, S3) based on their temporal order, where S1 contains four days for a better warm-up, while S2 and S3 contain three days. An additional 100,000 samples are collected on the following day after each stage for evaluation. The observed performance improvements across these three stages could provide strong empirical evidence of the effectiveness of research. 

An example of our sequence data is presented in Table~\ref{tab:rec_search_samples}. Each sequence is identified by the universally unique identifier of each page view (i.e., recommendation list presented to users). It mainly contains three fields: \textbf{\romannumeral1}) \textit{action\_seq} represents the sequence of historical interactions occurring before the current page view. To ensure efficient inference and compatibility with online deployment, we limit the maximum length of each sequence to 100 and truncate those longer sequences. Items within each sequence are represented by item\_ids and separated by commas. \textbf{\romannumeral2}) \textit{target\_item} refers to the items interacted with on the current page. For training, only the first clicked item is treated as the target. In contrast, all interacting items are considered as targets to compute metrics more comprehensively. \textbf{\romannumeral3}) \textit{query} is None for recommendation, while set to the corresponding query keywords for the search task, which are also collected from the real user inputs.

\textbf{\romannumeral2)} {Item Info}. Following previous benchmarks~\citep{yuan2022tenrec,liurecflow}, each item in \emph{Seq Data} is represented by a single-token identifier (item\_id). The mapping between item\_ids in the action\_seq and their corresponding SIDs is detailed in the Item Info of Table~\ref{tab:item_info}. To standardize the generation of SIDs, in addition to the constructed SIDs of each item $i$ (\texttt{base} and \texttt{Final}), FORGE enriches each item ID with additional multimodal information $\mathcal{H}^i$ derived by both the image and text of item $i$, serving as input for the encoding processes.
To preserve privacy, we release the processed \textit{multimodal embeddings} along with product titles processed via Named Entity Recognition (NER). 
Additionally, items $i^{+}$ \textit{with the most collaborative relations} of each item $i$ (i2i for simplicity) are provided as a field of each item and aid in SID generation in Section~\ref{feature extraction}. 

\subsubsection{Comparison with Other Datasets}
As depicted in Table~\ref{tab:dataset_stats}, AL-GR distinguishes itself from existing recommendation benchmarks in two aspects. First, early datasets such as MovieLens-20M~\citep{harper2015movielens} and Yelp~\citep{asghar2016yelp} are relatively small in scale, with limited user-item interactions, which makes them unsuitable for modeling large-scale traffic typically in industrial systems. Subsequent efforts~\citep{gao2022kuairec,liurecflow} alleviate this with abundant interactions from real-world platforms. In Table~\ref{tab:dataset_stats}, AL-GR further significantly extends the largest dataset, TenRec~\citep{yuan2022tenrec}, by approximately 100 times, providing over 14 billion user behaviors. Training and testing across multiple time periods also guarantee its reliability. Second, AL-GR is the first dataset to support SID construction at industrial scale. Unlike previous datasets that typically include only one or two modalities, AL-GR integrates three complementary modalities across billions of items to enable better SID generation. Moreover, the inclusion of query attributes allows it to be used for evaluating search tasks.

\subsection{Evaluation Metrics}
\label{Evaluation Metrics}
In this paper, we adopt the recommendation performance metric \textbf{HitRate} (HR) at the \textbf{item level} to evaluate the quality of generated SIDs. Taking a 3-level SID with 8192 codebooks per level as an example, \textbf{\romannumeral1}) FORGE first generates a list of 1200 candidate SIDs based on the beam search of GR. \textbf{\romannumeral2}) We map these SIDs back to their corresponding original items. \textbf{\romannumeral3}) The HitRate@K is measured by selecting the top-K \textbf{items} with the highest scores and checking whether each of them matches the ground truth items in the session.

However, evaluation with HR requires full GR training, which is typically time-consuming. To address this, FORGE introduces \textbf{Embedding HitRate} and \textbf{Gini coefficient} to measure the SID quality directly. Specifically, \textbf{\romannumeral1)} Embedding HitRate regards the multi-modal features $\mathcal{H}^i$ as the semantic alternatives of traditional item embeddings. It is calculated with a separate item-to-item retrieval~\cite{yang2020large} of all historical data based on the similarity (e.g., inner product) among the multi-modal features $\mathcal{H}^i$ to provide a preliminary evaluation of $\mathcal{H}^i$. \textbf{\romannumeral2)} The Gini coefficient measures the fairness of SIDs. Let $\mathcal{S}=\{s_1,\ldots,s_{N_d}\}$ denote the set of all possible SIDs (e.g., $C_18191C_28191C_38191$), where $N_d=8192^3$ for a 3-level SID with 8192 codewords per level. For each SID $s_j$, let $a_j$ be the number of items assigned to it. We sort these counts in non-decreasing order as $a_{(1)} \leq \cdots \leq
a_{(N_d)}$. The Gini coefficient is then computed as:
\begin{align}
    &G = \frac{2}{N_d}\sum_{i=1}^{N_d}
    \left(\frac{i}{N_d}-L(i)\right), \\
    &L(i) = \frac{C(i)}{C(N_d)}, \\
    &C(i) = \sum_{j=1}^{i} a_{(j)} .
\end{align}
Here, $C(i)$ denotes the cumulative number of items assigned to the first $i$ sorted SIDs, and $L(i)$ is the corresponding Lorenz curve. The Gini coefficient measures the deviation between a uniform SID distribution and the actual item distribution, where a lower value indicates a more balanced assignment. Our experiments show that both metrics are strongly correlated with HR, thus enabling efficient SID evaluation without additional GR training.

%% file: files/experiment.tex
\section{Experiments}
In this section, we conduct experiments on both the Amazon dataset and our proposed AL-GR datasets with three consecutive stages (S1, S2, and S3). For validation, we sample data from the day following each stage. By default, we employ RQ-VAE for SID construction and Qwen2.5-0.5B for generative retrieval. The goal of these experiments is to address the following research questions:
\begin{itemize}[left=0pt, itemindent=0pt, align=left]
    \item \textbf{RQ1:} Which SID construction strategies are beneficial for downstream generative retrieval in terms of HitRate?
    \item \textbf{RQ2:} How effectively can our proposed metrics assess SID quality without full GR training?
    \item \textbf{RQ3:} How is the generalization ability of the SID construction-based insights validated in our benchmark?
\end{itemize}

\begin{table}[htb]
\centering
\vspace{-0.2cm}
\caption{The effectiveness of SIDs configurations compared with \texttt{base} on our released AL-GR dataset across 3 stages. \texttt{Base} is constructed using limited item-to-item collaborative information defined in Equation~\ref{contrastive}, along with a KNN-based post-processing with threshold $\sigma$ set to 25, while \texttt{Final} integrates \texttt{sideinfo}, \texttt{i2i} and \texttt{Random-5} together for SID generation.}
\vspace{-0.2cm}
\label{tab:ablation_full}
\setlength{\tabcolsep}{2.5pt}  
\renewcommand{\arraystretch}{1.1}
\resizebox{1.\columnwidth}{!}{
    \begin{tabular}{l l r r r r r r r r}
    \toprule
    Stage & Method & \multicolumn{2}{c}{HR@20} & \multicolumn{2}{c}{HR@100} & \multicolumn{2}{c}{HR@500} & \multicolumn{2}{c}{HR@1000} \\
    \midrule
    \multirow{7}{*}{S1}
        & base      
            & 3.61\% & {\color{lightgray}\small +0.00\%} 
            & 9.67\% & {\color{lightgray}\small +0.00\%} 
            & 20.82\% & {\color{lightgray}\small +0.00\%} 
            & 25.09\% & {\color{lightgray}\small +0.00\%} \\
        & +KNN-10                           
            & 3.91\% & {\color{lightgray}\small +8.31\%} 
            & 10.32\% & {\color{lightgray}\small +6.72\%} 
            & 21.85\% & {\color{lightgray}\small +4.95\%} 
            & 25.38\% & {\color{lightgray}\small +1.16\%} \\
        & +KNN-5                            
            & 4.18\% & {\color{lightgray}\small +15.79\%} 
            & 10.81\% & {\color{lightgray}\small +11.79\%} 
            & 22.14\% & {\color{lightgray}\small +6.34\%} 
            & 25.30\% & {\color{lightgray}\small +0.84\%} \\
        & +Random-5                
            & 4.67\% & {\color{lightgray}\small +29.36\%} 
            & 11.79\% & {\color{lightgray}\small +21.92\%} 
            & 23.65\% & {\color{lightgray}\small +13.59\%} 
            & 26.30\% & {\color{lightgray}\small +4.82\%} \\
        & +i2i                       
            & 3.79\% & {\color{lightgray}\small +4.99\%} 
            & 10.04\% & {\color{lightgray}\small +3.83\%} 
            & 21.71\% & {\color{lightgray}\small +4.27\%} 
            & 26.97\% & {\color{lightgray}\small +7.49\%} \\
        & +sideinfo                       
            & 3.81\% & {\color{lightgray}\small +5.54\%} 
            & 10.00\% & {\color{lightgray}\small +3.41\%} 
            & 21.69\% & {\color{lightgray}\small +4.18\%} 
            & 26.88\% & {\color{lightgray}\small +7.13\%} \\
        & Final          
            & \textbf{4.89\%} & {\color{lightgray}\small \textbf{+35.46\%}} 
            & \textbf{12.31\%} & {\color{lightgray}\small \textbf{+27.30\%}} 
            & \textbf{24.79\%} & {\color{lightgray}\small \textbf{+19.07\%}} 
            & \textbf{29.01\%} & {\color{lightgray}\small \textbf{+15.62\%}} \\
    \midrule
    \multirow{7}{*}{S2}
        & base   
            & 4.15\% & {\color{lightgray}\small +0.00\%} 
            & 10.70\% & {\color{lightgray}\small +0.00\%} 
            & 22.71\% & {\color{lightgray}\small +0.00\%} 
            & 26.70\% & {\color{lightgray}\small +0.00\%} \\
        & +KNN-10                           
            & 4.63\% & {\color{lightgray}\small +11.57\%} 
            & 11.52\% & {\color{lightgray}\small +7.66\%} 
            & 23.24\% & {\color{lightgray}\small +2.33\%} 
            & 26.80\% & {\color{lightgray}\small +0.37\%} \\
        & +KNN-5                            
            & 4.81\% & {\color{lightgray}\small +15.90\%} 
            & 11.90\% & {\color{lightgray}\small +11.21\%} 
            & 24.18\% & {\color{lightgray}\small +6.47\%} 
            & 27.09\% & {\color{lightgray}\small +1.46\%} \\
        & +Random-5                
            & 5.23\% & {\color{lightgray}\small +26.02\%} 
            & 12.88\% & {\color{lightgray}\small +20.37\%} 
            & 25.05\% & {\color{lightgray}\small +10.30\%} 
            & 27.74\% & {\color{lightgray}\small +3.90\%} \\
        & +i2i                       
            & 4.12\% & {\color{mygreen}\small -0.72\%} 
            & 10.69\% & {\color{mygreen}\small -0.09\%} 
            & 23.06\% & {\color{lightgray}\small +1.54\%} 
            & 28.10\% & {\color{lightgray}\small +5.24\%} \\
        & +sideinfo                       
            & 4.32\% & {\color{lightgray}\small +4.10\%} 
            & 11.06\% & {\color{lightgray}\small +3.36\%} 
            & 23.46\% & {\color{lightgray}\small +3.30\%} 
            & 28.43\% & {\color{lightgray}\small +6.48\%} \\
        & Final           
            & \textbf{5.33\%} & {\color{lightgray}\small \textbf{+28.43\%}} 
            & \textbf{13.23\%} & {\color{lightgray}\small \textbf{+23.64\%}} 
            & \textbf{26.37\%} & {\color{lightgray}\small \textbf{+16.12\%}} 
            & \textbf{30.28\%} & {\color{lightgray}\small \textbf{+13.41\%}} \\
    \midrule
    \multirow{7}{*}{S3}
        & base        
            & 4.33\% & {\color{lightgray}\small +0.00\%} 
            & 11.16\% & {\color{lightgray}\small +0.00\%} 
            & 23.26\% & {\color{lightgray}\small +0.00\%} 
            & 27.24\% & {\color{lightgray}\small +0.00\%} \\
        & +KNN-10                           
            & 4.56\% & {\color{lightgray}\small +5.31\%} 
            & 11.78\% & {\color{lightgray}\small +5.56\%} 
            & 24.06\% & {\color{lightgray}\small +3.44\%} 
            & 27.54\% & {\color{lightgray}\small +1.10\%} \\
        & +KNN-5                            
            & 5.08\% & {\color{lightgray}\small +17.32\%} 
            & 12.61\% & {\color{lightgray}\small +12.99\%} 
            & 24.66\% & {\color{lightgray}\small +6.02\%} 
            & 27.72\% & {\color{lightgray}\small +1.76\%} \\
        & +Random-5                
            & 5.43\% & {\color{lightgray}\small +25.40\%} 
            & 13.14\% & {\color{lightgray}\small +17.74\%} 
            & 25.39\% & {\color{lightgray}\small +9.16\%} 
            & 28.03\% & {\color{lightgray}\small +2.90\%} \\
        & +i2i                       
            & 4.22\% & {\color{mygreen}\small -2.54\%} 
            & 11.32\% & {\color{lightgray}\small +1.43\%} 
            & 23.69\% & {\color{lightgray}\small +1.85\%} 
            & 28.91\% & {\color{lightgray}\small +6.13\%} \\
        & +sideinfo                       
            & 4.34\% & {\color{lightgray}\small +0.23\%} 
            & 11.43\% & {\color{lightgray}\small +2.42\%} 
            & 24.20\% & {\color{lightgray}\small +4.04\%} 
            & 29.32\% & {\color{lightgray}\small +7.64\%} \\
        & Final           
            & \textbf{5.44\%} & {\color{lightgray}\small \textbf{+25.64\%}} 
            & \textbf{13.79\%} & {\color{lightgray}\small \textbf{+23.57\%}} 
            & \textbf{26.71\%} & {\color{lightgray}\small \textbf{+14.83\%}} 
            & \textbf{30.86\%} & {\color{lightgray}\small \textbf{+13.29\%}} \\
    \bottomrule
    \vspace{-0.8cm}
    \end{tabular}
}
\end{table}

\subsection{Insights into SID Construction (RQ1)}
\label{RQ1}

\subsubsection{Overall Analysis}
Table~\ref{tab:ablation_Amazon} and Table~\ref{tab:ablation_full} present the ablation results based on the 3-level codewords with 8192 codebooks per level (i.e., 3$\times$8192), separately. From the results, we can conclude the following findings from two perspectives: 
\textbf{\romannumeral1)} \textbf{Post-processing to reduce collisions after SID generation leads to a fairer SID distribution and significantly improves retrieval performance.} \texttt{+KNN-5} and \texttt{+KNN-10} impose a limit on the number of items each SID can represent using the KNN-based solution. The improvements from \texttt{base} to \texttt{+KNN-10} and \texttt{+KNN-5} confirm the benefit of increasing fairness and representation granularity. In contrast, \texttt{+Random-5} randomly assigns only the last codeword of each SID to an item, irrespective of semantic similarity. This method yields a more balanced SID distribution and further boosts retrieval performance, suggesting that the final layer of SIDs is less critical than maintaining high utilization or fairness across the SID space.
\textbf{\romannumeral2)} \textbf{Incorporating higher-quality modalities enhances the precision of SID construction.} \texttt{+i2i} introduces co-occurrence-based item relationships into the contrastive learning of Equation~\ref{contrastive}, while \texttt{+sideinfo} enriches the SID generation with additional textual item metadata (e.g., categories, sellers) by concatenating the embeddings with the original text embedding. Although \texttt{+i2i} slightly underperforms the \texttt{base} model in terms of HR@20, both strategies outperform the baseline at higher thresholds (e.g., HR@1000), achieving over 5\% improvements across all training stages. This highlights the value of richer input signals in enhancing the effectiveness of GR.
By integrating all these strategies, \texttt{Final} using more \texttt{sideinfo} and \texttt{i2i} data with \texttt{Random-5} for SID generation demonstrates consistently more than 13\% increase compared to \texttt{base}.

\begin{table}[htb]
\centering
\vspace{-0.2cm}
\caption{The effectiveness of SIDs configurations compared with \texttt{base} on the Amazon dataset.}
\vspace{-0.2cm}
\label{tab:ablation_Amazon}
\setlength{\tabcolsep}{3pt}
\renewcommand{\arraystretch}{1.1}
\resizebox{1.\columnwidth}{!}{
    \begin{tabular}{l r r r r r r r r}
    \toprule
    Method & \multicolumn{2}{c}{HR@20} & \multicolumn{2}{c}{HR@100} & \multicolumn{2}{c}{HR@500} & \multicolumn{2}{c}{HR@1000} \\
    \midrule
         base      
            & 1.02\% & {\color{lightgray}\small +0.00\%} 
            & 2.47\% & {\color{lightgray}\small +0.00\%} 
            & 5.71\% & {\color{lightgray}\small +0.00\%} 
            & 8.07\% & {\color{lightgray}\small +0.00\%} \\
        noco      
            & 0.96\% & {\color{mygreen}\small -5.88\%} 
            & 1.83\% & {\color{mygreen}\small -25.91\%} 
            & 3.56\% & {\color{mygreen}\small -37.65\%} 
            & 4.97\% & {\color{mygreen}\small -38.41\%} \\
        +KNN-10                           
            & 1.60\% & {\color{lightgray}\small +56.86\%} 
            & 3.30\% & {\color{lightgray}\small +33.60\%} 
            & 7.19\% & {\color{lightgray}\small +25.92\%} 
            & 9.87\% & {\color{lightgray}\small +22.30\%} \\
        +KNN-5                            
            & 1.74\% & {\color{lightgray}\small +70.59\%} 
            & 3.80\% & {\color{lightgray}\small +53.85\%} 
            & 8.35\% & {\color{lightgray}\small +46.23\%} 
            & 11.28\% & {\color{lightgray}\small +39.78\%} \\
        +Random-5                
            & 2.05\% & {\color{lightgray}\small +100.98\%} 
            & 4.18\% & {\color{lightgray}\small +69.23\%} 
            & 9.32\% & {\color{lightgray}\small +63.22\%} 
            & 12.21\% & {\color{lightgray}\small +51.30\%} \\
        +sideinfo                       
            & 1.62\% & {\color{lightgray}\small +58.82\%} 
            & 3.19\% & {\color{lightgray}\small +29.15\%} 
            & 6.82\% & {\color{lightgray}\small +19.44\%} 
            & 9.41\% & {\color{lightgray}\small +16.60\%} \\
        Final          
            & \textbf{2.12\%} & {\color{lightgray}\small \textbf{+107.84\%}} 
            & \textbf{4.56\%} & {\color{lightgray}\small \textbf{+84.62\%}} 
            & \textbf{9.79\%} & {\color{lightgray}\small \textbf{+71.45\%}} 
            & \textbf{13.04\%} & {\color{lightgray}\small \textbf{+61.59\%}}\\
    \bottomrule
    \vspace{-0.6cm}
    \end{tabular}
}
\end{table}

\subsubsection{Analysis of different relations among items}
In this section, we explore how different relations can influence the SID generation in Equation~\ref{contrastive} and subsequent retrieval performance. We could conclude from the result in Table~\ref{tab:ablation_custom_v2} that i2i (collaborative items) relationships in recommendation systems can provide additional benefits for SID generation. Both the query2i and samecate (items within the same category) approaches underperform the \texttt{base} method, which utilizes only limited i2i data for alignment as in Equation~\ref{contrastive}. Besides, Table~\ref{tab:ablation_custom_v2} also demonstrates that incorporating item-side information (such as seller, price, and category) along with richer i2i relations could lead to more meaningful SIDs and improved retrieval performance.

\begin{table}[htb]
\centering
\caption{Relative improvements of different data modalities over the \texttt{base} model (3$\times$8192). Improvements are computed as $(M - \text{base}) / \text{base}$.}
\vspace{-0.15cm}
\label{tab:ablation_custom_v2}
\setlength{\tabcolsep}{5pt}
\renewcommand{\arraystretch}{1.1}
\resizebox{\columnwidth}{!}{
\begin{tabular}{l l r r r r r r r r}
\toprule
Stage & Method & \multicolumn{2}{c}{HR@20} & \multicolumn{2}{c}{HR@100} & \multicolumn{2}{c}{HR@500} & \multicolumn{2}{c}{HR@1000} \\
\midrule
\multirow{5}{*}{S1}
    & base      
        & 3.61\% & {\color{lightgray}\small +0.00\%} 
        & 9.67\% & {\color{lightgray}\small +0.00\%} 
        & 20.82\% & {\color{lightgray}\small +0.00\%} 
        & 25.09\% & {\color{lightgray}\small +0.00\%} \\
    & query2i       
        & 3.20\% & {\color{mygreen}\small -11.36\%} 
        & 8.70\% & {\color{mygreen}\small -10.03\%} 
        & 19.01\% & {\color{mygreen}\small -8.69\%} 
        & 23.93\% & {\color{mygreen}\small -4.62\%} \\
    & samecate  
        & 3.44\% & {\color{mygreen}\small -4.71\%} 
        & 8.92\% & {\color{mygreen}\small -7.76\%} 
        & 19.43\% & {\color{mygreen}\small -6.68\%} 
        & 23.94\% & {\color{mygreen}\small -4.58\%} \\
    & +sideinfo  
        & 3.81\% & {\color{lightgray}\small +5.54\%} 
        & 10.00\% & {\color{lightgray}\small +3.41\%} 
        & 21.69\% & {\color{lightgray}\small +4.18\%} 
        & 26.88\% & {\color{lightgray}\small +7.13\%} \\
    & +i2i    
        & 3.79\% & {\color{lightgray}\small +4.99\%} 
        & 10.04\% & {\color{lightgray}\small +3.83\%} 
        & 21.71\% & {\color{lightgray}\small +4.27\%} 
        & 26.97\% & {\color{lightgray}\small +7.49\%} \\
\midrule
\multirow{5}{*}{S2}
    & base      
        & 4.15\% & {\color{lightgray}\small +0.00\%} 
        & 10.70\% & {\color{lightgray}\small +0.00\%} 
        & 22.71\% & {\color{lightgray}\small +0.00\%} 
        & 26.70\% & {\color{lightgray}\small +0.00\%} \\
    & query2i       
        & 3.71\% & {\color{mygreen}\small -10.60\%} 
        & 9.69\% & {\color{mygreen}\small -9.44\%} 
        & 20.67\% & {\color{mygreen}\small -8.98\%} 
        & 25.72\% & {\color{mygreen}\small -3.67\%} \\
    & samecate  
        & 3.81\% & {\color{mygreen}\small -8.19\%} 
        & 9.86\% & {\color{mygreen}\small -7.85\%} 
        & 21.31\% & {\color{mygreen}\small -6.16\%} 
        & 25.78\% & {\color{mygreen}\small -3.45\%} \\
    & +sideinfo  
        & 4.32\% & {\color{lightgray}\small +4.10\%} 
        & 11.06\% & {\color{lightgray}\small +3.36\%} 
        & 23.46\% & {\color{lightgray}\small +3.30\%} 
        & 28.43\% & {\color{lightgray}\small +6.48\%} \\
    & +i2i    
        & 4.12\% & {\color{mygreen}\small -0.72\%} 
        & 10.69\% & {\color{mygreen}\small -0.09\%} 
        & 23.06\% & {\color{lightgray}\small +1.54\%} 
        & 28.10\% & {\color{lightgray}\small +5.24\%} \\
\midrule
\multirow{5}{*}{S3}
    & base      
        & 4.33\% & {\color{lightgray}\small +0.00\%} 
        & 11.16\% & {\color{lightgray}\small +0.00\%} 
        & 23.26\% & {\color{lightgray}\small +0.00\%} 
        & 27.24\% & {\color{lightgray}\small +0.00\%} \\
    & query2i       
        & 3.67\% & {\color{mygreen}\small -15.24\%} 
        & 9.99\% & {\color{mygreen}\small -10.48\%} 
        & 21.77\% & {\color{mygreen}\small -6.41\%} 
        & 27.05\% & {\color{mygreen}\small -0.70\%} \\
    & samecate  
        & 3.98\% & {\color{mygreen}\small -8.08\%} 
        & 10.23\% & {\color{mygreen}\small -8.33\%} 
        & 21.91\% & {\color{mygreen}\small -5.80\%} 
        & 26.50\% & {\color{mygreen}\small -2.72\%} \\
    & +sideinfo  
        & 4.34\% & {\color{lightgray}\small +0.23\%} 
        & 11.43\% & {\color{lightgray}\small +2.42\%} 
        & 24.20\% & {\color{lightgray}\small +4.04\%} 
        & 29.32\% & {\color{lightgray}\small +7.64\%} \\
    & +i2i    
        & 4.22\% & {\color{mygreen}\small -2.54\%} 
        & 11.32\% & {\color{lightgray}\small +1.43\%} 
        & 23.69\% & {\color{lightgray}\small +1.85\%} 
        & 28.91\% & {\color{lightgray}\small +6.13\%} \\
\bottomrule
\vspace{-1.0cm}
\end{tabular}
}
\end{table}

\subsubsection{Ablation Study of the Multimodal feature}
Table~\ref{tab:data_modality_ablation} presents how different modalities influence the final representation $\mathcal{H}^i$ through the offline experiment in the first stage of AL-GR. The results indicate that using merely text and image for representation generation leads to a significant performance drop. In comparison, the i2i-based method is more effective for recommendation tasks, but it still underperforms compared to the fusion of multiple modalities, especially when it comes to HR@1000. These results indicate that each of the three modalities contributes meaningfully to the construction of high-quality SIDs, thereby improving retrieval performance.

\begin{table}[hbt]
\centering
\vspace{-0.10cm}
\caption{Ablation study about the influence of the data modality during SID generation.}
\vspace{-0.15cm}

\label{tab:data_modality_ablation}
\renewcommand{\arraystretch}{1}
\resizebox{0.9 \columnwidth}{!}{
\begin{tabular}{l l r r r r}
\toprule
Stage & Data Modality & HR@20 & HR@100 & HR@500 & HR@1000 \\
\midrule
\multirow{4}{*}{S1}
    & text-only   & 3.01\% & 7.97\% & 18.00\% & 21.01\% \\
    & image-only  & 3.06\% & 8.34\%& 18.79\% & 22.03\% \\
    & i2i-only  & 4.63\% & 11.58\% & 22.52\% & 23.15\% \\
    & Final & \textbf{4.89\%} & \textbf{12.31\%} & \textbf{24.79\%} & \textbf{29.01\%} \\
\bottomrule
\vspace{-0.7cm}
\end{tabular}
}
\end{table}

\begin{table}[htb]
\centering
\caption{Relative improvements of different ID collision alleviating strategies over the \texttt{base} model (3$\times$8192).}
\vspace{-0.30cm}
\label{tab:ablation_custom}
\setlength{\tabcolsep}{4pt}  
\renewcommand{\arraystretch}{1.1}
\resizebox{\columnwidth}{!}{
\begin{tabular}{l l r r r r r r r r}
\toprule
Stage & Method & \multicolumn{2}{c}{HR@20} & \multicolumn{2}{c}{HR@100} & \multicolumn{2}{c}{HR@500} & \multicolumn{2}{c}{HR@1000} \\
\midrule
\multirow{6}{*}{S1}
    & base      
        & 3.61\% & {\color{lightgray}\small +0.00\%} 
        & 9.67\% & {\color{lightgray}\small +0.00\%} 
        & 20.82\% & {\color{lightgray}\small +0.00\%} 
        & 25.09\% & {\color{lightgray}\small +0.00\%} \\
    & noco      
        & 2.96\% & {\color{mygreen}\small -18.01\%} 
        & 7.69\% & {\color{mygreen}\small -20.48\%} 
        & 17.63\% & {\color{mygreen}\small -15.32\%} 
        & 22.27\% & {\color{mygreen}\small -11.24\%} \\
    & merge     
        & 3.00\% & {\color{mygreen}\small -16.90\%} 
        & 8.32\% & {\color{mygreen}\small -13.96\%} 
        & 18.66\% & {\color{mygreen}\small -10.37\%} 
        & 24.22\% & {\color{mygreen}\small -3.47\%} \\
    & +KNN-10     
        & 3.91\% & {\color{lightgray}\small +8.31\%} 
        & 10.32\% & {\color{lightgray}\small +6.72\%} 
        & 21.85\% & {\color{lightgray}\small +4.95\%} 
        & 25.38\% & {\color{lightgray}\small +1.16\%} \\
    & +KNN-5      
        & 4.18\% & {\color{lightgray}\small +15.79\%} 
        & 10.81\% & {\color{lightgray}\small +11.79\%} 
        & 22.14\% & {\color{lightgray}\small +6.34\%} 
        & 25.30\% & {\color{lightgray}\small +0.84\%} \\
    & +Random-5 
        & 4.67\% & {\color{lightgray}\small +29.36\%} 
        & 11.79\% & {\color{lightgray}\small +21.92\%} 
        & 23.65\% & {\color{lightgray}\small +13.59\%} 
        & 26.30\% & {\color{lightgray}\small +4.82\%} \\
\midrule
\multirow{6}{*}{S2}
    & base      
        & 4.15\% & {\color{lightgray}\small +0.00\%} 
        & 10.70\% & {\color{lightgray}\small +0.00\%} 
        & 22.71\% & {\color{lightgray}\small +0.00\%} 
        & 26.70\% & {\color{lightgray}\small +0.00\%} \\
    & noco      
        & 3.49\% & {\color{mygreen}\small -15.90\%} 
        & 8.91\% & {\color{mygreen}\small -16.73\%} 
        & 19.47\% & {\color{mygreen}\small -14.27\%} 
        & 24.07\% & {\color{mygreen}\small -9.85\%} \\
    & merge     
        & 3.54\% & {\color{mygreen}\small -14.70\%} 
        & 9.11\% & {\color{mygreen}\small -14.86\%} 
        & 20.34\% & {\color{mygreen}\small -10.44\%} 
        & 26.06\% & {\color{mygreen}\small -2.40\%} \\
    & +KNN-10     
        & 4.63\% & {\color{lightgray}\small +11.57\%} 
        & 11.52\% & {\color{lightgray}\small +7.66\%} 
        & 23.24\% & {\color{lightgray}\small +2.33\%} 
        & 26.80\% & {\color{lightgray}\small +0.37\%} \\
    & +KNN-5      
        & 4.81\% & {\color{lightgray}\small +15.90\%} 
        & 11.90\% & {\color{lightgray}\small +11.21\%} 
        & 24.18\% & {\color{lightgray}\small +6.47\%} 
        & 27.09\% & {\color{lightgray}\small +1.46\%} \\
    & +Random-5
        & 5.23\% & {\color{lightgray}\small +26.02\%} 
        & 12.88\% & {\color{lightgray}\small +20.37\%} 
        & 25.05\% & {\color{lightgray}\small +10.30\%} 
        & 27.74\% & {\color{lightgray}\small +3.90\%} \\
\midrule
\multirow{6}{*}{S3}
    & base      
        & 4.33\% & {\color{lightgray}\small +0.00\%} 
        & 11.16\% & {\color{lightgray}\small +0.00\%} 
        & 23.26\% & {\color{lightgray}\small +0.00\%} 
        & 27.24\% & {\color{lightgray}\small +0.00\%} \\
    & noco      
        & 3.59\% & {\color{mygreen}\small -17.09\%} 
        & 9.24\% & {\color{mygreen}\small -17.20\%} 
        & 20.29\% & {\color{mygreen}\small -12.77\%} 
        & 24.84\% & {\color{mygreen}\small -8.81\%} \\
    & merge     
        & 3.56\% & {\color{mygreen}\small -17.78\%} 
        & 9.55\% & {\color{mygreen}\small -14.43\%} 
        & 20.65\% & {\color{mygreen}\small -11.22\%} 
        & 26.86\% & {\color{mygreen}\small -1.40\%} \\
    & +KNN-10     
        & 4.56\% & {\color{lightgray}\small +5.31\%} 
        & 11.78\% & {\color{lightgray}\small +5.56\%} 
        & 24.06\% & {\color{lightgray}\small +3.44\%} 
        & 27.54\% & {\color{lightgray}\small +1.10\%} \\
    & +KNN-5     
        & 5.08\% & {\color{lightgray}\small +17.32\%} 
        & 12.61\% & {\color{lightgray}\small +12.99\%} 
        & 24.66\% & {\color{lightgray}\small +6.02\%} 
        & 27.72\% & {\color{lightgray}\small +1.76\%} \\
    & +Random-5 
        & 5.43\% & {\color{lightgray}\small +25.40\%} 
        & 13.14\% & {\color{lightgray}\small +17.74\%} 
        & 25.39\% & {\color{lightgray}\small +9.16\%} 
        & 28.03\% & {\color{lightgray}\small +2.90\%} \\
\bottomrule
\vspace{-1.1cm}
\end{tabular}
}
\end{table}

\subsubsection{Analysis of the SID post-processing}
Table~\ref{tab:ablation_custom} clarifies the impact of different ID collision strategies on the GR models. Compared to the \texttt{base} method, which employs a KNN-based strategy to limit the number of items per SID to fewer than 25, the \texttt{noco} variant without any post-processing shows significantly reduced effectiveness. Furthermore, results also indicate that more effective collision avoidance (\texttt{noco}$\rightarrow$\texttt{base}$\rightarrow$\texttt{KNN-10}$\rightarrow$\texttt{KNN-5}) contributes to higher semantic encoding fairness and leads to higher metrics. 

To further validate this, we conduct an inverse experiment using the \texttt{merge} method, which intentionally degrades the SID fairness by merging adjacent SIDs below a certain threshold. The substantial drop in performance compared to \texttt{base} confirms a strong positive correlation between SID utilization and retrieval accuracy.

Unlike the KNN-based policy, \texttt{Random-5} leverages a random-based approach to assign the last level codeword sequentially and randomly. While this reduces collisions and maintains high codebook utilization and SID fairness, it might sacrifice meaningful semantic information at the final quantization level. However, the continuous HitRate gain suggests that utilization and collision are more important than the semantic features, as features in the last level capture limited information in the residual quantization.

\begin{table}[htb]
\centering
\caption{Relative improvements of different SID level structures over the \texttt{base} model (3$\times$8192).}
\vspace{-0.2cm}
\label{tab:ablation_codebook}
\setlength{\tabcolsep}{3pt}
\renewcommand{\arraystretch}{1.1}
\resizebox{\columnwidth}{!}{
\begin{tabular}{l l r r r r r r r r}
\toprule
Stage & Method & \multicolumn{2}{c}{HR@20} & \multicolumn{2}{c}{HR@100} & \multicolumn{2}{c}{HR@500} & \multicolumn{2}{c}{HR@1000} \\
\midrule
\multirow{7}{*}{S1}
    & base              
        & 3.61\% & {\color{lightgray}\small +0.00\%} 
        & 9.67\% & {\color{lightgray}\small +0.00\%} 
        & 20.82\% & {\color{lightgray}\small +0.00\%} 
        & 25.09\% & {\color{lightgray}\small +0.00\%} \\
    & 2048\_4096\_8192  
        & 3.20\% & {\color{mygreen}\small -11.36\%} 
        & 8.70\% & {\color{mygreen}\small -10.03\%} 
        & 19.01\% & {\color{mygreen}\small -8.69\%} 
        & 23.93\% & {\color{mygreen}\small -4.62\%} \\
    & 8192\_4096\_2048  
        & 3.44\% & {\color{mygreen}\small -4.71\%} 
        & 8.92\% & {\color{mygreen}\small -7.76\%} 
        & 19.43\% & {\color{mygreen}\small -6.68\%} 
        & 23.94\% & {\color{mygreen}\small -4.58\%} \\
    & 2x32768           
        & 3.37\% & {\color{mygreen}\small -6.65\%} 
        & 9.09\% & {\color{mygreen}\small -6.00\%} 
        & 20.33\% & {\color{mygreen}\small -2.35\%} 
        & 26.94\% & {\color{lightgray}\small +7.37\%} \\
    & 3x512             
        & 2.84\% & {\color{mygreen}\small -21.33\%} 
        & 7.80\% & {\color{mygreen}\small -19.34\%} 
        & 17.63\% & {\color{mygreen}\small -15.32\%} 
        & 23.33\% & {\color{mygreen}\small -7.01\%} \\
    & 1024\_4096\_32768 
        & 3.76\% & {\color{lightgray}\small +4.16\%} 
        & 10.09\% & {\color{lightgray}\small +4.34\%} 
        & 22.03\% & {\color{lightgray}\small +5.81\%} 
        & 27.95\% & {\color{lightgray}\small +11.40\%} \\
    & 4x4096            
        & 3.65\% & {\color{lightgray}\small +1.11\%} 
        & 9.50\% & {\color{mygreen}\small -1.76\%} 
        & 20.25\% & {\color{mygreen}\small -2.74\%} 
        & 25.06\% & {\color{mygreen}\small -0.12\%} \\
\midrule
\multirow{7}{*}{S2}
    & base              
        & 4.15\% & {\color{lightgray}\small +0.00\%} 
        & 10.70\% & {\color{lightgray}\small +0.00\%} 
        & 22.71\% & {\color{lightgray}\small +0.00\%} 
        & 26.70\% & {\color{lightgray}\small +0.00\%} \\
    & 2048\_4096\_8192  
        & 3.96\% & {\color{mygreen}\small -4.58\%} 
        & 10.44\% & {\color{mygreen}\small -2.43\%} 
        & 22.17\% & {\color{mygreen}\small -2.38\%} 
        & 27.64\% & {\color{lightgray}\small +3.52\%} \\
    & 8192\_4096\_2048  
        & 4.02\% & {\color{mygreen}\small -3.13\%} 
        & 10.52\% & {\color{mygreen}\small -1.68\%} 
        & 22.54\% & {\color{mygreen}\small -0.75\%} 
        & 28.21\% & {\color{lightgray}\small +5.66\%} \\
    & 2x32768           
        & 3.85\% & {\color{mygreen}\small -7.23\%} 
        & 10.03\% & {\color{mygreen}\small -6.26\%} 
        & 21.93\% & {\color{mygreen}\small -3.43\%} 
        & 28.96\% & {\color{lightgray}\small +8.46\%} \\
    & 3x512             
        & 3.34\% & {\color{mygreen}\small -19.52\%} 
        & 8.65\% & {\color{mygreen}\small -19.16\%} 
        & 19.46\% & {\color{mygreen}\small -14.31\%} 
        & 24.89\% & {\color{mygreen}\small -6.78\%} \\
    & 1024\_4096\_32768 
        & 4.34\% & {\color{lightgray}\small +4.58\%} 
        & 11.07\% & {\color{lightgray}\small +3.46\%} 
        & 23.58\% & {\color{lightgray}\small +3.83\%} 
        & 29.75\% & {\color{lightgray}\small +11.42\%} \\
    & 4x4096            
        & 4.19\% & {\color{lightgray}\small +0.96\%} 
        & 10.53\% & {\color{mygreen}\small -1.59\%} 
        & 21.49\% & {\color{mygreen}\small -5.37\%} 
        & 26.11\% & {\color{mygreen}\small -2.21\%} \\
\midrule
\multirow{7}{*}{S3}
    & base              
        & 4.33\% & {\color{lightgray}\small +0.00\%} 
        & 11.16\% & {\color{lightgray}\small +0.00\%} 
        & 23.26\% & {\color{lightgray}\small +0.00\%} 
        & 27.24\% & {\color{lightgray}\small +0.00\%} \\
    & 2048\_4096\_8192  
        & 4.04\% & {\color{mygreen}\small -6.70\%} 
        & 10.50\% & {\color{mygreen}\small -5.91\%} 
        & 23.15\% & {\color{mygreen}\small -0.47\%} 
        & 29.07\% & {\color{lightgray}\small +6.72\%} \\
    & 8192\_4096\_2048  
        & 4.19\% & {\color{mygreen}\small -3.23\%} 
        & 10.92\% & {\color{mygreen}\small -2.15\%} 
        & 23.59\% & {\color{lightgray}\small +1.42\%} 
        & 28.99\% & {\color{lightgray}\small +6.42\%} \\
    & 2x32768           
        & 3.91\% & {\color{mygreen}\small -9.70\%} 
        & 10.49\% & {\color{mygreen}\small -6.00\%} 
        & 22.65\% & {\color{mygreen}\small -2.62\%} 
        & 30.00\% & {\color{lightgray}\small +10.13\%} \\
    & 3x512             
        & 3.49\% & {\color{mygreen}\small -19.40\%} 
        & 9.07\% & {\color{mygreen}\small -18.73\%} 
        & 19.99\% & {\color{mygreen}\small -14.06\%} 
        & 25.96\% & {\color{mygreen}\small -4.70\%} \\
    & 1024\_4096\_32768 
        & 4.30\% & {\color{mygreen}\small -0.69\%} 
        & 11.17\% & {\color{lightgray}\small +0.09\%} 
        & 24.14\% & {\color{lightgray}\small +3.78\%} 
        & 30.35\% & {\color{lightgray}\small +11.42\%} \\
    & 4x4096            
        & 4.00\% & {\color{mygreen}\small -7.62\%} 
        & 10.46\% & {\color{mygreen}\small -6.27\%} 
        & 21.72\% & {\color{mygreen}\small -6.62\%} 
        & 26.45\% & {\color{mygreen}\small -2.90\%} \\
\bottomrule
\vspace{-1.0cm}
\end{tabular}
}
\end{table}

\subsubsection{Analysis of the SID Structure}
In previous experiments, we primarily considered SID structures such as 3$\times$8192. This raises the question of \texttt{whether alternative structures may offer improved performance in GRs}, such as those with more codewords (e.g., 4$\times$4096) or distinct hierarchy (e.g., 2048\_4096\_8192). Different structures may change both the capacity and the decoding cost of SIDs, making it necessary to examine their effects systematically. Extensive validation presented in Table~\ref{tab:ablation_codebook} indicates that 3-level SIDs still represent the optimal trade-off between efficiency and effectiveness. In contrast, 2-level structures exhibit significantly worse performance, while the 4-level structures incur higher inference costs with negligible gains. Although the 1024\_4096\_32768 configuration achieves a 10\% improvement in HR@1000 compared to \texttt{base} in Stage 1, the results in Stage 3 also indicate that these improvements gradually diminish as the model converges. Therefore, once the item space represented by SIDs reaches a certain level, the structural characteristics have limited impacts on GRs.

\begin{figure*}[hbt]
\centering
\begin{subfigure}[t]{0.4\linewidth}
  \centering
  \includegraphics[width=\linewidth]{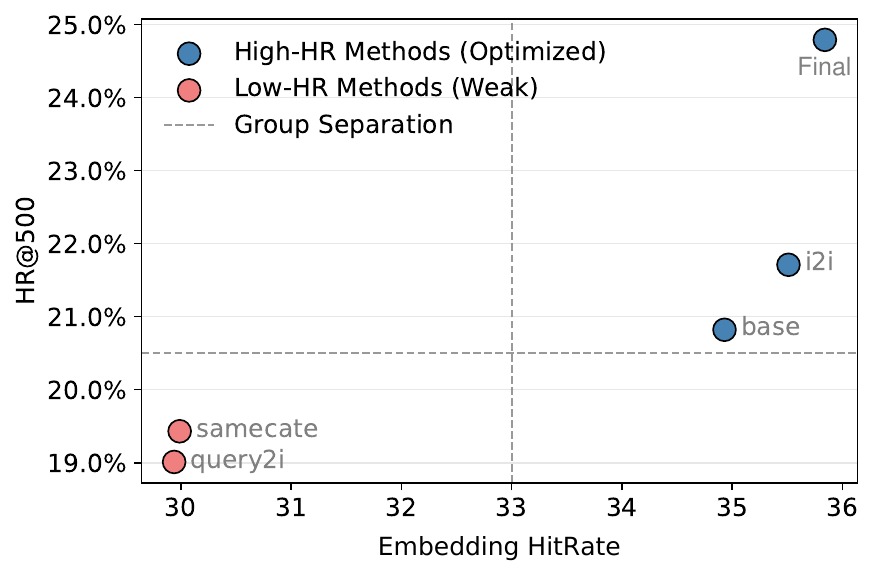}
  \label{fig:gul_hr}
\end{subfigure}
\begin{subfigure}[t]{0.4\linewidth}
  \centering
  \includegraphics[width=\linewidth]{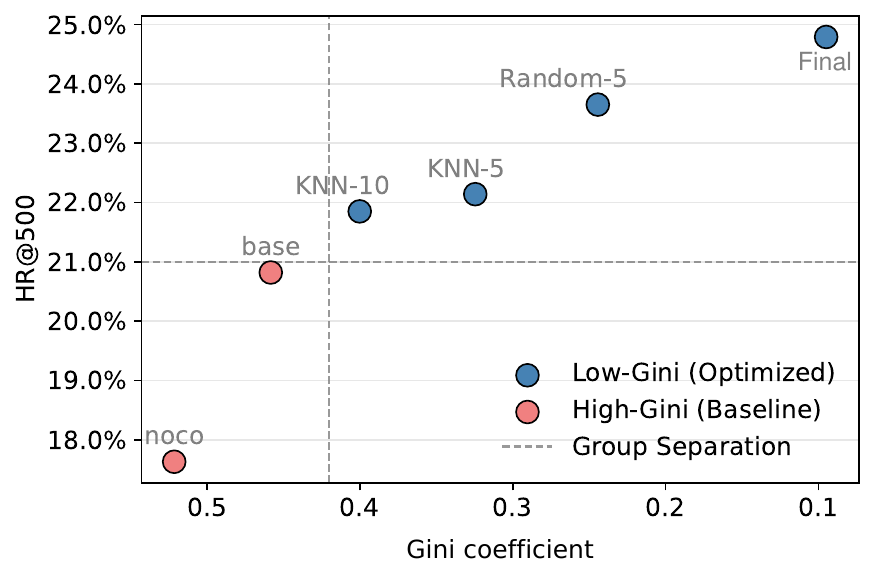}
  \label{fig:gini_hr}
\end{subfigure}
\vspace{-0.7cm}
\caption{
The effectiveness of Embedding HitRate and Gini coefficient for SID evaluation.
  Left: Positive correlation between Embedding HitRate and retrieval performance.
  Right: Lower Gini coefficient (fairer SID usage) leads to higher HR@500.
}
\vspace{-0.36cm}
\label{fig:correlation_combined}
\end{figure*}

\subsubsection{Analysis of the SID Encoding}
As other related works~\cite{deng2025onerec,zhou2025onerec,chen2025onesearch} have explored alternative encoding methods such as RQ-Kmeans, we also conduct a comparative analysis and present the results in Table~\ref{tab:quantization_comparison}. First, both the semantic-related and residual decoding strategies play a critical role in retrieval performance. The \texttt{Random} baseline assigns SIDs to items without considering any multimodal features, leading to a significant 70\% drop in HR@1000. In contrast, the \texttt{Multiple-VQ} approach employs separate encoders at each level, resulting in decoupled codewords within SIDs and consequently suboptimal representation learning. Second, \texttt{RQ-Kmeans} and \texttt{OPQ} achieve performance comparable to \texttt{RQ-VAE} (base). As noted by OneRec~\cite{deng2025onerec}, a key advantage of RQ-Kmeans lies in its efficient use of codebooks. However, through the insight of appropriate post-processing to handle ID collisions, \texttt{Final} implementation with \texttt{RQ-VAE} demonstrates a clear superiority over \texttt{RQ-Kmeans}.

\begin{table}[htb]
\centering
\vspace{-0.26cm}
\caption{Comparison of different encoding methods.}
\vspace{-0.2cm}
\label{tab:quantization_comparison}
\setlength{\tabcolsep}{8pt}  
\renewcommand{\arraystretch}{1.2}  
\resizebox{\columnwidth}{!}{
\begin{tabular}{c c c c c c}
\toprule
Stage & Method & HR@20 & HR@100 & HR@500 & HR@1000 \\
\midrule
\multirow{6}{*}{S1}
    & Random     & 1.08\% & 2.91\% & 4.82\% & 4.82\% \\
    & Multiple-VQ         & 2.87\% & 8.15\% & 18.30\% & 23.84\% \\
    & RQ-Kmeans  & 3.42\% & 9.39\% & 20.71\% & \underline{26.11\%} \\
    & OPQ & 3.57\% & 9.57\% & 21.19\% & 26.64\% \\
    & RQ-VAE (base)     & \underline{3.61\%} & \underline{9.67\%} & \underline{20.82\%} & 25.09\% \\
    & RQ-VAE (Final)     & \textbf{4.89\%} & \textbf{12.31\%} & \textbf{24.79\%} & \textbf{29.01\%} \\
\midrule
\multirow{5}{*}{S2}
    & Random     & 1.66\% & 4.09\% & 6.66\% & 6.66\% \\
    & Multiple-VQ         & 3.33\% & 8.68\% & 19.55\% & 25.36\% \\
    & RQ-Kmeans  & 3.79\% & 10.09\% & 22.23\% & \underline{27.95\%} \\
    & RQ-VAE (base)     & \underline{4.15\%} & \underline{10.70\%} & \underline{22.71\%} & 26.70\% \\
    & RQ-VAE (Final)     & \textbf{5.33\%} & \textbf{13.23\%} & \textbf{26.37\%} & \textbf{30.28\%} \\
    
\midrule
\multirow{5}{*}{S3}
    & Random     & 1.51\% & 4.04\% & 6.41\% & 6.41\% \\
    & Multiple-VQ         & 3.47\% & 9.28\% & 20.69\% & 26.47\% \\
    & RQ-Kmeans  & 3.96\% & 10.82\% & 23.00\% & \underline{28.71\%} \\
    & RQ-VAE (base)     & \underline{4.33\%} & \underline{11.16\%} & \underline{23.26\%} & 27.24\% \\
    & RQ-VAE (Final)     & \textbf{5.44\%} & \textbf{13.79\%} & \textbf{26.71\%} & \textbf{30.86\%} \\
\bottomrule
\end{tabular}
}
\vspace{-0.5cm}
\end{table}

\subsection{Direct Metrics for the SID Evaluation (RQ2)}

The direct evaluations of SIDs using Embedding HitRate and Gini coefficient are presented in Figure~\ref{fig:correlation_combined}. As we mentioned in Section~\ref{Evaluation Metrics}, the Embedding HitRate captures the quality of item collaborative relationships within the multi-modal feature $\mathcal{H}^i$, while the Gini coefficient reflects the fairness of SID distribution. \textbf{First}, it is evident that leveraging more related information could lead to the improvement of the associated metrics. 
From Figure~\ref{fig:correlation_combined}, we could observe that incorporating richer relational information consistently improves these metrics. For instance, \texttt{i2i} with more collaborative item-item interactions achieves a higher Embedding HitRate compared to \texttt{base}.
Likewise, a more refined SID collision strategy could lead to higher codebook fairness and thus a lower Gini coefficient (e.g., \texttt{noco}$\rightarrow$\texttt{base}$\rightarrow$\texttt{KNN-10}$\rightarrow$\texttt{KNN-5}$\rightarrow$\texttt{Random-5}).
We \textbf{then} train the retrieval model with all these SIDs across three stages, and find that both metrics have a strong relation with the ultimate HitRate. These results suggest that \textbf{it is not strictly necessary to train recommenders} to assess the quality of SIDs. Instead, our proposed metrics provide a reliable and efficient proxy for evaluating SID effectiveness, offering a practical and scalable solution for future research and applications.

\subsection{Analysis of the Generalizability of SID Optimization Insights (RQ3)}

In this part, we conduct additional experiments to investigate whether the empirical insights from Section~\ref{RQ1} are robust across different configurations, architectures, and tasks.

\begin{table}[hb]
\vspace{-0.25cm}
\centering
\caption{Performance of \texttt{base} and \texttt{Final} under $2\times32768$ level.}
\vspace{-0.3cm}
\label{tab:codebook_2x32768_comparison}
\setlength{\tabcolsep}{1.5pt}  
\resizebox{0.78\columnwidth}{!}{
    \begin{tabular}{c c r r r r}
    \toprule
    Stage & Version & HR@20 & HR@100 & HR@500 & HR@1000 \\
    \midrule
    \multirow{2}{*}{S1}
        & base    & 3.37\% & 9.09\% & 20.33\% & 26.94\% \\
        & Final   & \textbf{4.57\%} & \textbf{11.94\%} & \textbf{24.67\%} & \textbf{31.34\%} \\
    \midrule
    \multirow{2}{*}{S2}
        & base    & 3.85\% & 10.03\% & 21.93\% & 28.96\% \\
        & Final   & \textbf{5.08\%} & \textbf{12.61\%} & \textbf{26.17\%} & \textbf{33.53\%} \\
    \midrule
    \multirow{2}{*}{S3}
        & base    & 3.91\% & 10.49\% & 22.65\% & 30.00\% \\
        & Final   & \textbf{5.11\%} & \textbf{13.04\%} & \textbf{26.88\%} & \textbf{34.15\%} \\
    \bottomrule
    \vspace{-0.8cm}
    \end{tabular}
}
\end{table}

\subsubsection{Generalization to the SID Level.} 
The comparison between \texttt{base} and \texttt{Final} using 2-level SIDs
(i.e., 32768 codebooks each level) is placed in Table~\ref{tab:codebook_2x32768_comparison}. This setup maintains the total representational capacity of the original multi-level design but distributes it across fewer levels. The improvements from \texttt{base} to \texttt{Final} verify that our optimization generalizes well to configurations with fewer levels. Across three stages, \texttt{Final} achieves more than a 10\% increase in HR@100 and HR@1000 over \texttt{base}.

\begin{table}[hbt]\small
\centering
\vspace{-0.25cm}
\caption{Retrieval performance on the generative search task of Taobao across stages.}
\vspace{-0.2cm}
\label{tab:generative_search_ablation}
\renewcommand{\arraystretch}{1}
\begin{tabular}{l l r r r r}
\toprule
Stage & Codebook & HR@20 & HR@100 & HR@500 & HR@1000 \\
\midrule
\multirow{4}{*}{S1}
    & 3$\times$8192 (base)   & 14.30\% & 27.91\% & 43.95\% & 50.74\% \\
    & 3$\times$8192 (Final)  & \textbf{23.39\%} & \textbf{37.05\%} & \textbf{51.77\%} & \textbf{56.89\%} \\
    & 2$\times$32768 (base)  & 11.53\% & 24.09\% & 39.65\% & 47.21\% \\
    & 2$\times$32768 (Final) & \textbf{17.77\%} & \textbf{31.56\%} & \textbf{46.98\%} & \textbf{53.91\%} \\
\midrule
\multirow{4}{*}{S2}
    & 3$\times$8192 (base)   & 14.93\% & 28.98\% & 46.21\% & 53.10\% \\
    & 3$\times$8192 (Final)  & \textbf{24.20\%} & \textbf{38.76\%} & \textbf{54.08\%} & \textbf{59.36\%} \\
    & 2$\times$32768 (base)  & 12.14\% & 25.38\% & 41.77\% & 49.47\% \\
    & 2$\times$32768 (Final) & \textbf{18.89\%} & \textbf{32.98\%} & \textbf{49.14\%} & \textbf{56.10\%} \\
\midrule
\multirow{4}{*}{S3}
    & 3$\times$8192 (base)   & 15.08\% & 29.83\% & 47.83\% & 55.10\% \\
    & 3$\times$8192 (Final)  & \textbf{24.71\%} & \textbf{40.06\%} & \textbf{56.04\%} & \textbf{61.46\%} \\
    & 2$\times$32768 (base)  & 12.54\% & 26.26\% & 43.48\% & 51.57\% \\
    & 2$\times$32768 (Final) & \textbf{19.13\%} & \textbf{34.19\%} & \textbf{51.18\%} & \textbf{58.34\%} \\
\bottomrule
\vspace{-0.8cm}
\end{tabular}
\end{table}

\subsubsection{Generalization to the Search Domain.}
In industrial scenarios, mobile apps often involve multiple tasks such as search and recommendation. Building a separate SID for each task can be resource-intensive. In this section, we investigate whether SIDs trained on the recommendation task can generalize to the search task. Related query keywords used in this section are collected from the search tab of Taobao and released in our AL-GR dataset. The input for the search task presented in Table~\ref{tab:search_format} is similar to the recommendation, except for the query keywords of each user. As shown in Table~\ref{tab:generative_search_ablation}, our findings in the search domain align closely with those observed in recommendation, where the 3-level SID consistently outperforms the 2-layer SID. Moreover, across all configurations, the proposed SID optimization strategy demonstrates clear superiority over the baseline. This phenomenon underscores its potential for enabling versatile SIDs in real-world scenarios.

\begin{table*}[htb]
\centering
\caption{Performance of \texttt{base} and \texttt{Final} under different architectures and model sizes in Stage S1.}
\vspace{-0.2cm}
\label{tab:ablation_t5_qwen}
\setlength{\tabcolsep}{2.0pt}
\renewcommand{\arraystretch}{1.1}
\resizebox{0.68\linewidth}{!}{
    \begin{tabular}{l *{8}{r}}
    \toprule
    \multirow{2}{*}{Codebook} 
      & \multicolumn{4}{c}{\texttt{t5-base (0.2B)}} 
      & \multicolumn{4}{c}{\texttt{Qwen2.5-3B}} \\
    \cmidrule(lr){2-5} \cmidrule(lr){6-9}
      & HR@20 & HR@100 & HR@500 & HR@1000 
      & HR@20 & HR@100 & HR@500 & HR@1000 \\
    \midrule
    3$\times$8192 (base)   
      & 1.29\% & 3.62\% & 8.91\%  & 11.98\% 
      & 4.49\% & 11.71\% & 24.40\% & 28.94\% \\
    3$\times$8192 (Final)  
      & \textbf{2.36\%} & \textbf{5.99\%} & \textbf{12.98\%} & \textbf{16.08\%}
      & \textbf{5.60\%} & \textbf{14.18\%} & \textbf{27.58\%} & \textbf{32.12\%} \\
    \midrule
    2$\times$32768 (base)  
      & 1.84\% & 5.14\% & 11.98\% & 16.45\%
      & 3.72\% & 10.19\% & 22.64\% & 29.75\% \\
    2$\times$32768 (Final) 
      & \textbf{2.46\%} & \textbf{6.55\%} & \textbf{14.30\%} & \textbf{18.88\%}
      & \textbf{4.96\%} & \textbf{12.80\%} & \textbf{27.13\%} & \textbf{34.29\%} \\
    \bottomrule
    \vspace{-0.7cm}
    \end{tabular}
}
\end{table*}

\subsubsection{Influence of the Model Size.} The findings of FORGE can also be extended to larger models with 3B parameters, shown in the right part of Table~\ref{tab:ablation_t5_qwen}, where the SID configurations bring about 4\% improvements in HR@1000.
Moreover, increasing the model size consistently enhances retrieval performance, leading to nearly 10\% improvements across all settings compared to the \texttt{Qwen2.5-0.5B} baseline in Table~\ref{tab:ablation_full} and~\ref{tab:codebook_2x32768_comparison}. We attribute this to the greater knowledge retention capacity enabled by the expanded parameter count.

\subsubsection{Generalization to the GR Architecture.}
The left panel of Table~\ref{tab:ablation_t5_qwen} presents the performance of \texttt{t5-base}~\cite{raffel2020exploring} with an encoder-decoder architecture. While it differs from the decoder-only setup used in previous experiments, the insights of SID optimizations verified in this paper still yield substantial improvements across different architectural paradigms. The consistent performance gain over the \texttt{base} model for both types of SIDs (i.e., $2\times 32768$ and $3\times 8192$) highlights our broad applicability.

\begin{figure}[htbp]
    \centering
    \vspace{-0.12cm}
    \includegraphics[width=0.35\textwidth]{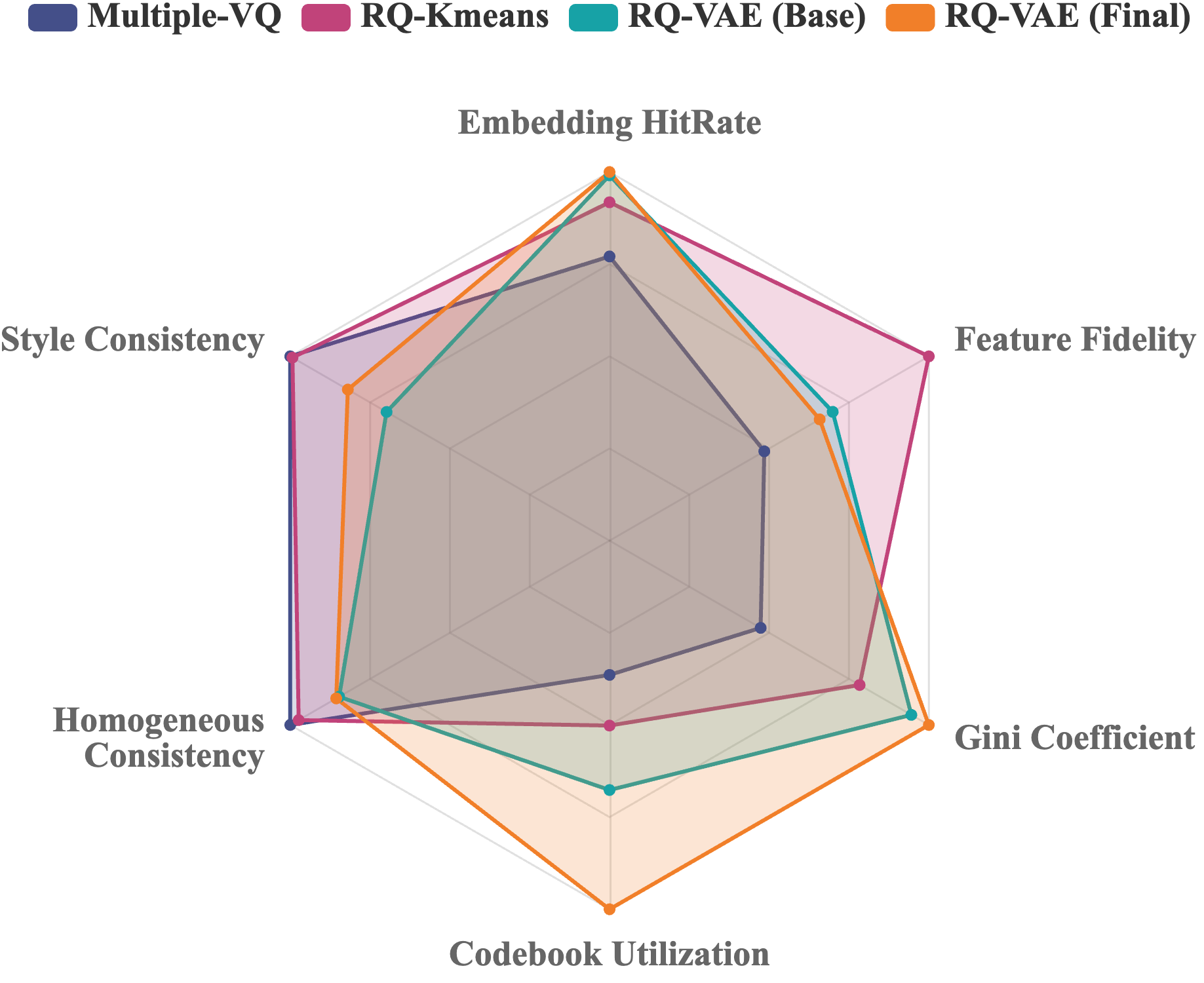}
    \vspace{-0.35cm}  
    \caption{Performance comparison under different metrics.}
    \vspace{-0.45cm} 
    \label{fig:additonal_metrics}
\end{figure}

\subsubsection{Additional Evaluation Metrics.} 
Apart from the Embedding HitRate and Gini coefficient discussed above, we further introduce several additional metrics to examine their relationships with downstream GR performance: \textbf{\romannumeral 1)} Feature Fidelity measures how well the original multimodal semantics are preserved after RQ-VAE quantization. \textbf{\romannumeral 2)} Style Consistency evaluates whether items with the same style are assigned to the same SID. \textbf{\romannumeral 3)} Homogeneous Consistency evaluates whether items derived from the same source are assigned to the same SID. \textbf{\romannumeral 4)} Codebook Utilization measures the proportion of codebooks that are actually used to represent items.

We have two main observations. \textbf{\romannumeral 1)} As discussed previously, Embedding HitRate, fairness, and codebook utilization are positively correlated with retrieval performance. The consistent ranking, i.e., \texttt{Final} $>$ \texttt{base} $>$ \texttt{RQ-Kmeans} $>$ \texttt{Multiple-VQ}, observed in both Figure~\ref{fig:additonal_metrics} and Table~\ref{tab:quantization_comparison}, further validates this conclusion. \textbf{\romannumeral 2)} In contrast, other metrics such as feature fidelity do not show a clear correlation with retrieval effectiveness. For example, \texttt{Multiple-VQ} achieves the best scores on two of these metrics, yet obtains the worst retrieval results in Table~\ref{tab:quantization_comparison}. A similar trend can be observed for \texttt{RQ-Kmeans}, which significantly outperforms \texttt{base} in feature fidelity but still lags in overall retrieval performance.

\begin{table}[htbp]
\centering
\vspace{+0.08cm}
\caption{Results of the 7-day online A/B study with our SID insights deployed in the "Guess You Like" section of Taobao.}
\vspace{-0.2cm}
\label{tab:online_study}
\setlength{\tabcolsep}{4pt}  
\renewcommand{\arraystretch}{1}
\resizebox{0.66\linewidth}{!}{
    \begin{tabular}{c c c}
    \toprule
    PVR & HitRate & Transaction Count \\
    \midrule
    \textbf{+8.93\%} & \textbf{+10.02\%} & \textbf{+0.35\%} \\
    \bottomrule
    \end{tabular}
}
\vspace{-0.35cm}
\end{table}

\subsubsection{Online Study.} To enable a more realistic and comprehensive evaluation pipeline, we have deployed the model on the homepage of Taobao for recommendations, where the downstream ranking stage could further refine the retrieval items before presenting them to users. During serving, when a user clicks on an item detail page, an asynchronous inference request is triggered. The RTP-LLM inference framework~\cite{rtp-llm} would start the retrieval process and write output candidates into the cache table in 200ms, ensuring no impact on the main recommendation service's performance.

The improvement through 7 days of online experiments is presented in Table~\ref{tab:online_study} to serve as strong evidence, where our configurations improve the PVR (Product View Rate) and HR by 8.93\% and 10.02\% compared to \texttt{base}. Moreover, replacing \texttt{base} with \texttt{Final} ultimately results in a 0.35\% absolute increase in user transactions.

%% file: files/conclusion.tex
\section{Conclusion}
In this paper, we propose \textbf{FORGE}, a comprehensive benchmark for forming semantic identifiers in generative
retrieval in industrial datasets. We begin by systematically exploring significant yet less-explored configurations of SIDs and investigating several strategies from multiple angles. To validate their influence on downstream GRs, we carry out extensive comparisons and in-depth analysis on both research and industrial datasets. To enhance the reliability of our experimental findings, we verify these insights on different downstream GR models, SID structures, and tasks. All the results demonstrate the generalization of findings in this paper.
Additionally, FORGE introduces two metrics for efficiently evaluating SID quality without requiring expensive GR training. 
To facilitate the application and reproducibility, we have released the industrial dataset AL-GR used in this paper, which contains 14 billion interactions and 251 million items collected from Taobao. We hope that FORGE and AL-GR will serve as a valuable resource for the community, encouraging further innovation in this area.

%% file: files/appendix.tex
\appendix

\section{Implementation Details}
\label{App: Implementation Details}
\subsection{Running environment.} The experiments are conducted on the PPU 810E, a
GPU architecture developed by Alibaba. This platform achieves approximately 90\% of
the computational performance of the NVIDIA A100 GPU, while featuring a high-capacity
memory system with 95.6 GB of on-board memory.

\subsection{Hyperparameter configurations.} Table~\ref{tab:config_summary} and~\ref{tab:training_configurations} summarize the hyperparameters and configuration of models in the SID generation task and generative retrieval task, respectively. For the ID generation task, RQ-VAE employs symmetric 3-layer MLPs (512-64-512) with a 64D discrete codebook and CN-CLIP fusion. The training lasts for 150 epochs using AdamW for stable convergence. For the generative retrieval task, we set the maximum lengths of the historical behaviors to 100, and employ two sequence-to-sequence models in our experiments: a \texttt{T5-Base} model and a \texttt{Qwen2.5-0.5B} model. Both models are fine-tuned on the same dataset with a maximum token length of 1280, where the maximum source and target lengths are set to 1024 and 256, respectively. The T5-Base model is trained for 4 epochs with a per-device batch size of 80, resulting in a total batch size of 1280 using 16 PPU-ZW810E GPUs. It uses a constant learning rate of $2 \times 10^{-4}$
and the AdamW optimizer with default parameters. In contrast, the Qwen2.5-0.5B-Instruct model is trained for a single epoch with a smaller per-device batch size of 40, also achieving the same total batch size of 1280 using 32 PPU-ZW810E GPUs. It adopts a linear learning rate scheduler with a base learning rate of $5 \times 10^{-5}$. Both models are trained using bfloat16 precision to improve training efficiency. The \texttt{Qwen2.5-3B} model follows the same training configuration as the \texttt{Qwen2.5-0.5B} model. All the SIDs are regarded as new tokens and included in the tokenizer of LLMs upon initialization. The dynamic beam size of 3-level and 2-level SIDs during inference is set to \{300, 600, 1200\} and \{600, 1200\}, respectively.

\begin{table}[htb]
\centering
\vspace{-0.2cm}
\caption{Hyperparameters and training configurations of the two models used in the Generative Retrival Task.}
\vspace{-0.3cm}
\label{tab:training_configurations}
\resizebox{\columnwidth}{!}{
\begin{tabular}{lcc}
\toprule
 & \textbf{T5-Base} & \textbf{Qwen2.5-0.5B-Instruct} \\
\midrule
Seq Length & 100 & 100 \\
Model Type & T5 & Qwen2.5 \\
Checkpoint & google-t5/t5-base & Qwen/Qwen2.5-0.5B-Instruct \\
Max Length & 1280 & 1280 \\
Max Source Length & 1024 & 1024 \\
Max Target Length & 256 & 256 \\
Epochs & 4 & 1 \\
Batch Size (per device) & 80 & 40 \\
Total Train Batch Size & 1280 & 1280 \\
Learning Rate & 2e-4 & 5e-5 \\
LR Scheduler & constant & linear \\
Optimizer& \multicolumn{2}{c}{AdamW (betas=[0.9, 0.999], eps=1e-8)}  \\
bf16 & \checkmark & \checkmark \\
GPU Count & 16 & 32 \\
GPU Type & PPU-ZW810E & PPU-ZW810E \\
Training Time & 20h/epoch& 80h/epoch \\
\bottomrule
\vspace{-0.9cm}
\end{tabular}
}
\end{table}

\begin{table}[ht]
\centering
\vspace{-0.15cm}
\caption{Core configuration of the RQ-VAE model and training setup.}
\vspace{-0.2cm}
\label{tab:config_summary}
\resizebox{\columnwidth}{!}{
\begin{tabular}{lll}
\toprule
 & \textbf{Parameter} & \textbf{Value} \\
\midrule
\multirow{6}{*}{Model} 
& Model Type & RQ-VAE (RQVAE\_EMBED\_CLIP) \\
& Codebook Dim & 64 \\
& Input Dim & 512 \\
& Encoder Structure & 3-LayerMLP (512-256-256-64, ReLU) \\
& Decoder Structure & 3-LayerMLP (64-256-256-512, ReLU) \\
& Fusion Structure & QFormer \\
& Multimodal Encoder & CN-Clip (chinese-clip-vit-base-patch16) \\
\midrule

\multirow{7}{*}{Training} 
& Epochs & 150 \\
& Warmup Epochs & 40 \\
& Learning Rate & 0.0002 \\
& Optimizer & AdamW (betas=[0.9, 0.999], eps=1e-8) \\
& LR Scheduler & cosine \\
& Batch Size & 2048 \\

\bottomrule
\vspace{-0.9cm}
\end{tabular}
}
\end{table}

\subsection{Instructions for GRs}
The instructions for the recommendation and search tasks are presented in Table~\ref{tab:recsys_format} and Table~\ref{tab:search_format}, respectively. Each sequence is represented as a series of semantic IDs (e.g., \texttt{C1220C8322C20452}). For a better LLM tokenization, for the $j$th-level codeword in our processed SID, we represent it with $c_j +\sum_{k=1}^{j-1}n_k$ (e.g., $C8193$ with $C_21$ as the 2nd codeword and 8192 codebooks each level). Each user behavior sequence is encoded as continuous SIDs with no separator. The inputs consist of both system and user instructions, while the answer is only made up of codewords of the target SID. Optionally, for the search task, the query keywords of each request are added to the user instruction and placed after the user sequence. The whole training process is to enable LLMs to understand the instructions and provide satisfying item lists given an input.

\begin{table*}[hbtp]
\centering
\caption{Example of our sequence data on recommendation and search tasks with segmented fields. For validation, all the interacted items during this page view are used as targets.}
\vspace{-0.3cm}
\label{tab:rec_search_samples}
\resizebox{0.82\textwidth}{!}{
\begin{tabular}{l c l c c c}
\toprule
\textbf{task} & \textbf{data split} & \textbf{target\_item} & \textbf{query} & \textbf{action\_seq} \\
\midrule

\multirow{2}{*}[-4ex]{Recsys} 
& train
& 835905354006
& /  & \makecell[l]{566476193770, 907995423965, 897259251809, \\916138422508, 841619863136, 841665058110,\\ 895390654577, 895382438483,  844457291108} \\
\addlinespace

& test
& \makecell[l]{817445445744, \\780409885810}
& /  & \makecell[l]{546901080156, 626466181497, 804805887937,\\ 819489082025, 919698707572, 680837256237,\\ 611110770249, 696394590725, 738397879056} \\
\addlinespace
\midrule
\multirow{2}{*}[-5ex]{Search} 
& train
& 762971121687 
& \makecell[l]{Fullerene Serum}
& \makecell[l]{937018558175, 936950943952, 947234504809, \\ 666112772468,  928130569440, 938702048112, \\940280943038, 918448888947,  792039516661} \\
\addlinespace

& test
& \makecell[l]{562133894238,\\567812038293,\\610069547271}
& Dried Mugwort Leaves 
& \makecell[l]{736070049413, 949041120166, 945601927485, \\ 922276261406, 801633068800, 901162945401,\\884983761715, 705197027339, 890974760840} \\
\bottomrule
\end{tabular}
}
\end{table*}

\begin{table*}[hbt]
\centering
\caption{Example of Item Info.}
\vspace{-0.2cm}
\label{tab:item_info}
\resizebox{0.75\textwidth}{!}{
\begin{tabular}{ccccc}
\toprule
\textbf{item\_id} & $\mathbf{\mathcal{H}}^i \in R^{512}$ & \textbf{semantic\_id} & \textbf{related\_item} & \textbf{title} \\
\toprule
835905354006 & [0.12, 0.56, ..., 0.03] & [1203,2315,3576] & 787551011877 & Women's blouse \\
\addlinespace
855036080309 & [0.04, 0.02, ..., 0.05] & [2706,4659,2176] & 545092516562 & Men's Plus-Size Pajamas \\
\bottomrule
\vspace{-0.5cm}
\end{tabular}
}
\end{table*}

\begin{table}[htb]
\centering
\caption{Example of the instruction-tuning dataset format for next-item prediction in e-commerce recommendation. The original data is in Chinese, and an English translation is provided for comprehension.}
\vspace{-0.15cm}
\label{tab:recsys_format}
\begin{tabular}{p{0.12\linewidth} p{0.8\linewidth}}
\toprule
\textbf{Field} & \textbf{Content} \\
\midrule
system & You are a recommendation system. Predict the user's next action in an e-commerce scenario based on their historical behavior. A sequence of past interactions in chronological order will be provided, with each behavior encoded as a single semantic ID. \\
\midrule
user & User's historical behavior sequence: \texttt{\seqsplit{C1220C8322C20452C6084C10195C20067C3256C14673C21112C7054C9412C18926C3021C10986C18869C3411C15297C23680C116C15928C19183C4405C11869C21320C7221C13888C16791C7743C16005C22091}}. Please predict the semantic ID of the user's next behavior in the e-commerce recommendation scenario. \\
\midrule
answer & \texttt{C3626C8758C22717} \\
\bottomrule
\end{tabular}
\end{table}

\section{Dataset}
The benefits of the proposed AL-GR can be classified into three aspects. \textbf{\romannumeral1) Semantic identifiers generation}: with the provided multi-modal features in Item-Info, users could construct their SIDs with different configurations (e.g., different SID levels, SID postprocessing, SID Encoding, etc.) and subsequently assess the quality of SIDs using our proposed two new metrics in Section~\ref{Evaluation Metrics}. \textbf{\romannumeral2) Generative recommendation/search with semantic identifiers}: researchers could further evaluate the effectiveness of their constructed SID with both recommendation and search data we provided in Seq Data. Moreover, they could also perform a thorough comparison between their proposed models and other SOTA baselines using our datasets. \textbf{\romannumeral3) Standard retrieval tasks using single token identifiers}: as we also release the original user interaction sequences in action\_seq using single token identifiers, it would be convenient for users to perform and conduct experiments on traditional retrieval with our data.

\begin{table}[htb]
\centering
\vspace{-0.2cm}
\caption{Example of the instruction-tuning dataset format for next-item prediction in e-commerce search.}
\vspace{-0.2cm}
\label{tab:search_format}
\begin{tabular}{p{0.12\linewidth} p{0.8\linewidth}}
\toprule
\textbf{Field} & \textbf{Content} \\
\midrule
system & You are a search system. Predict the user's next action in an e-commerce scenario based on their historical behavior and current query. The behavior sequence is given as semantic IDs, each representing a triplet of attributes, in chronological order. \\
\midrule
user & User's history: \texttt{\seqsplit{C5299C12905C22291C1477C12826C20935C1931C9821C24180C853C14294C22645C2075C12389C16385C853C13340C20935C4894C14014C18136C6398C13220C22786C6398C12675C23264C4707C8594C18742C6398C14580C20512C6398C9530C19761}}. Query of the: short cropped vest. Predict the semantic ID of the next behavior. \\
\midrule
answer & \texttt{C1832C15680C24359} \\
\bottomrule
\end{tabular}
\vspace{-0.5cm}
\end{table}